\documentclass[hyper,12pt,letterpaper]{JHEP3} 
\usepackage{latexsym,amsmath,amsfonts,amssymb}
\usepackage{graphicx}
\usepackage{bbm}
\usepackage{multirow}

\newcommand{\Nugual}[1]{\ensuremath{\mathcal{N}\!= #1}}

\newcommand{\parfrac}[2]{\frac{\partial #1}{\partial #2}}

\newcommand{\eg}{\textit{e.g.}}

\newcommand{\ie}{\textit{i.e.}}

\newcommand{\etc}{\textit{etc}}
\newcommand{\rep}[1]{\ensuremath{\mathbf{#1}}}
\newcommand{\mr}{\mathrm}

\numberwithin{equation}{section}

\newcommand{\be}{\begin{equation}} \newcommand{\ee}{\end{equation}}
\newcommand{\bea}{\begin{equation} \begin{aligned}} \newcommand{\eea}{\end{aligned} \end{equation}}
\newcommand{\tabs}{\rule[-1.2ex]{0pt}{3.5ex}}

\newcommand{\calD}{\mathcal{D}}

\newcommand{\calJ}{\mathcal{J}}

\newcommand{\calL}{\mathcal{L}}
\newcommand{\calM}{\mathcal{M}}

\newcommand{\calO}{\mathcal{O}}

\newcommand{\calT}{\mathcal{T}}
\newcommand{\calV}{\mathcal{V}}

\newcommand{\bbC}{\mathbb{C}}

\newcommand{\bbR}{\mathbb{R}}
\newcommand{\bbZ}{\mathbb{Z}}
\newcommand{\unit}{\mathbbm{1}}

\DeclareMathOperator{\Tr}{tr}

\def\cN{\mathcal{N}}
\def\cO{\mathcal{O}}

\def\bR{\mathbb{R}}
\def\bC{\mathbb{C}}
\def\SU{\mathrm{SU}}
\def\USp{\mathrm{USp}}
\def\OSp{\mathrm{OSp}}
\def\SO{\mathrm{SO}}
\def\SL{\mathrm{SL}}
\def\U{\mathrm{U}}
\def\tr{\mathop{\mathrm{tr}}}
\def\ch{\mathop{\mathrm{ch}}}
\def\vev#1{\langle#1\rangle}
\def\bX{\mathbb{X}}
\def\skipper{\hskip.5em\relax}
\def\Esix#1#2#3#4#5#6{%
{\text{\small$\begin{array}{c@{\skipper}c@{\skipper}c@{\skipper}c@{\skipper}c@{\skipper}c}
&&#2&& \\
#1 &  #3 &  #4 & #5 & #6
\end{array}$}}}
\def\cI{\mathcal{I}}
\def\adj{\ensuremath{\mathbf{adj}}}
\def\Sym{\mathop{\mathrm{Sym}}\nolimits}

\title{Sicilian gauge theories and \Nugual{1} dualities}

\let\AA\diamondsuit
\let\BB\heartsuit
\author{Francesco Benini$^\AA$, Yuji Tachikawa$^\BB$, and Brian Wecht$^\BB$\\

{\tt fbenini@princeton.edu, yujitach,bwecht@ias.edu}\\
$^\AA$ Department of Physics, Princeton University, \\
Princeton, NJ 08544, USA \\
$^\BB$ School of Natural Sciences, Institute for Advanced Study, \\
Princeton, NJ 08540, USA \\
}

\preprint{PUTP-2312}

\abstract{
In theories without known Lagrangian descriptions, knowledge of the global symmetries is often
one of the few pieces of information we have at our disposal. Gauging (part of) such
global symmetries can then lead to interesting new theories, which are usually still quite mysterious. In this work, we describe a set of tools that can be used to explore the superconformal phases of these theories. In particular, we describe the contribution of such non-Lagrangian sectors to
the NSVZ $\beta$-function, and elucidate the counting of marginal deformations.
We apply our techniques to \Nugual{1} theories obtained by mass deformations of the \Nugual{2} conformal theories recently found by Gaiotto. Because the basic building block of these theories is a triskelion, or trivalent vertex, we dub them ``Sicilian gauge theories.'' We identify these \Nugual{1}  theories as compactifications of the six-dimensional $A_N$ $(2,0)$ theory on Riemann surfaces with punctures
and $\SU(2)$ Wilson lines. These theories include the holographic duals of the \Nugual{1} supergravity solutions found by Maldacena and Nu\~nez.
}

\begin{document}
\section{Introduction}
%

Strongly coupled four-dimensional (4d) superconformal field theories (SCFTs)
have been intensely studied for many years, and there has been great progress in our understanding of their dynamics. The usual procedure
is to start with a weakly coupled Lagrangian description in the ultraviolet (UV), and to then try to find the properties of the theory in the infrared (IR) limit.  Since we usually require that the theories in question be asymptotically free, we typically know everything about the UV theory, although the low-energy phase often remains mysterious. In part because of their computational intractability, theories without known Lagrangian descriptions are comparatively rare,
and not well explored. Notable examples are
the $\cN=2$ theories with $E$-type flavor symmetry found by Minahan and Nemeschansky (MN) \cite{Minahan:1996fg,Minahan:1996cj}, and close relatives of them analyzed in  \cite{Ganor:1996xd,Gukov:1998kt}.

In retrospect, this situation strongly contrasts with the case in two dimensions: For 2d
CFTs a Lagrangian description is valuable when we have it,
but it is not an absolute necessity. Thanks to the infinite-dimensional symmetry algebra,
we can often understand a theory without recourse to a Lagrangian description.
In four dimensions the symmetry algebra is much smaller, so much less is known
about conformal field theories without weakly coupled Lagrangian descriptions. However, there is no reason {\it a priori} that theories with Lagrangian descriptions should dominate the space of all possible quantum field theories.
It is possible that by restricting our field of study to Lagrangian theories, we may be missing many strange beasts out in the wild.

Indeed, a recent work by Gaiotto \cite{Gaiotto:2009we} demonstrated
that there is an infinite family of \Nugual{2} non-Lagrangian theories, obtained by
compactifying six-dimensional \Nugual{(2,0)} SCFTs on a sphere with three superconformal defects.
The most important such theory is called $T_N$, which has flavor symmetry $\SU(N)^3$.
$T_2$ is just a collection of eight free chiral multiplets; $T_3$ is the $E_6$ theory of MN;
$T_N$ with $N>3$ are new SCFTs.
Gaiotto showed that these non-Lagrangian theories
arise naturally as essential building blocks of S-dual descriptions
of standard Lagrangian quiver gauge theories.
In the dual description, standard \Nugual{2} vector multiplets
couple to a subgroup of the flavor symmetry of the non-Lagrangian blocks.
Gaiotto's work put the pioneering observations by Argyres, Seiberg
and Wittig \cite{Argyres:2007cn,Argyres:2007tq} into a general framework, and as such greatly increased the set
of known non-Lagrangian SCFTs.
The gravity duals of these theories were identified in \cite{Gaiotto:2009gz},
and an alternative construction in terms of $(p,q)$ 5-branes and 7-branes was found in \cite{Benini:2009gi}. Wilson loops in these theories were discussed in \cite{Drukker:2009tz}, and other recent developments can be found in \eg{} \cite{Tachikawa:2009rb, Nanopoulos:2009xe, Maruyoshi:2009uk}.

In general, as in \cite{Argyres:2007cn,Argyres:2007tq}, Gaiotto's theories feature
a vector multiplet coupled to a non-Lagrangian theory.
Although this construction seemed strange when first used by Argyres and Seiberg, we now realize that
it is in fact a very common one.
Such non-Lagrangian theories can be used essentially in the same way as matter fields,
the only difference being that they remain strongly coupled in the limit where the gauge coupling goes to zero. One can ask what sorts of theories can be generated by using these non-Lagrangian theories as building blocks.

We can now greatly enlarge the space of supersymmetric field theories.
For example nothing stops us from coupling \Nugual{1} vector multiplets, instead of full \Nugual{2} multiplets, to the non-Lagrangian
blocks.  Indeed such theories naturally arise when we decouple
the adjoint chiral multiplets inside \Nugual{2} vector multiplets by giving them a mass.
In general, we can think of starting from an arbitrary combination of chiral multiplets and non-Lagrangian
field theories (the `matter'), and couple them by adding superpotential terms and gauging part of the
flavor symmetries.

The problem, then, is how to study such an enlarged class of theories.
When the gauge coupling is weak, it can be studied much as in General
Gauge Mediation \cite{Meade:2008wd}, by studying how current--current two-point functions
affect the gauge fields which couple to the flavor symmetry.
In principle the knowledge of higher-point functions of flavor currents would allow us to perform
the calculation as an expansion in powers of the gauge coupling.\footnote{In particular,
we would find how to consistently couple dynamical gauge fields to the flavor symmetry,
generating the analogue of sea-gull terms for charged scalars.}
In this paper we will skip this intricate stage of higher-loop calculations,
and instead directly move to the non-perturbative aspects.
This will turn out to be not as
difficult as it sounds, thanks to the many exact results available in SCFTs.

The central tool to be established is the extension to our situation
of the exact $\beta$-function
of Novikov-Shifman-Vainshtein-Zakharov (NSVZ) \cite{Novikov:1983uc,Novikov:1985rd},
which involves the anomalous dimensions of chiral elementary fields.
When the gauge group is coupled to a non-Lagrangian theory, we only have information about the chiral ring and current operators, and
not a description in terms of elementary fields.
We show that we can replace the terms involving anomalous dimensions with three-point functions which can be reliably calculated.
With the NSVZ $\beta$-function at our disposal, we will be able
to count the number of exactly marginal deformations of such theories
by generalizing the argument of Leigh and Strassler \cite{Leigh:1995ep}.

Our main application of these techniques will be the study of the \Nugual{1} theories
obtained by the mass deformations of adjoint scalars of the $\cN=2$
theories of Gaiotto \cite{Gaiotto:2009we}.
These theories have many copies of $T_N$ as the `matter content,' to whose flavor symmetry
many $\SU(N)$ gauge multiplets couple.
Gauge theories whose matter contents are bifundamentals are known as quiver gauge theories;
those containing copies of $T_N$ will be dubbed  ``Sicilian gauge theories,'' for reasons to be explained later.

We propose that these $\cN=1$ theories describe the IR physics
of the \Nugual{1} compactification of the
 6d $(2,0)$ theory on Riemann surfaces with punctures.
We identify their gravity duals as the solutions found by Maldacena and Nu\~nez \cite{Maldacena:2000mw}, and generalizations thereof (for example to include punctures). The main new feature is that $\SU(2)$ Wilson lines are allowed on the Riemann surface.
The parameter space is thus the space of Riemann surfaces with punctures and $\SU(2)$ Wilson lines (with fixed conjugacy class around the
punctures), on which the S-duality group naturally acts. This space is expected
to be the same as the conformal manifold $\calM_C$ of the field theory; we
provide various checks of this. This structure demonstrates the existence of an
intricate net of S-dualities between these \Nugual{1} theories (see \cite{Argyres:1999xu,Halmagyi:2004ju} for related work).

The structure of the paper is as follows.
We start in Sec.~\ref{sec: beta} by generalizing the NSVZ
$\beta$-function. This turns out to be straightforward, since by supersymmetry we can relate
the $\beta$-function to the anomaly of the superconformal R-symmetry. Since this is determined by a three-point function, we
can reliably compute it even when we do not know a Lagrangian description. We describe both the one-loop and all-orders $\beta$-functions, in the case
where we have both fundamental fields as well as a non-Lagrangian sector. A few examples are discussed.

In Sec.~\ref{sec: TN} we study $T_N$ theories coupled to \Nugual{1} $\SU(N)$ gauge multiplets.
We start by recalling well-known properties of the $T_N$ theory,
and then go on to count the number of exactly marginal deformations.
We identify the gravity duals of these theories as the \Nugual{1} solutions of Maldacena and Nu\~nez \cite{Maldacena:2000mw}.
We show the agreement of the central charges $a$ and $c$, the number of
exactly marginal deformations, and the dimension of the operator corresponding to a wrapped M2-brane.

We conclude with a short discussion in Sec.~\ref{sec: conclusion}.
There are many appendices:
In Appendix \ref{app: formulae} we collect standard fomul\ae{} for SCFTs;
in Appendix \ref{app: betafuncs} we give a more rigorous derivation of the $\beta$-function found in Section \ref{sec: beta};
in Appendix \ref{app: chiral ring relations} we study some chiral ring relations of $T_N$ required in our analysis;
in Appendix \ref{app: twisting} we recall the twistings of the 6d $(2,0)$ theory;
finally
in Appendix \ref{app: MN metric} we write down the explicit form of the solutions found by Maldacena and Nu\~nez.

%
\section{Exact $\beta$-functions}
\label{sec: beta}
%

A necessary condition for superconformality is that the $\beta$-function for the physical gauge coupling  $g_p$ vanishes. For $\cN=1$ supersymmetric theories, the general form of this $\beta$-function was found by NSVZ \cite{Novikov:1983uc} to be
\be
\label{nsvz}
\beta_{8\pi^2/g_p^2} \equiv \parfrac{}{\log M} \, \frac{8\pi^2}{g_p^2} = \frac{3T_2(\adj) - \sum_i T_2(\rep{r}_i) \big( 1 - \gamma_i(g_p) \big)}{1-\frac{g_p^2 \, T_2(\adj)}{8 \pi^2}} \;,
\ee
where $M$ is the energy scale, $\Tr T_{r_i}^a T_{r_i}^b = T_2(\rep{r}_i) \, \delta^{ab}$ gives the quadratic Casimir of the representation $\rep{r}_i$, and $\gamma_i(g_p)$ is the anomalous dimension of the matter field $\Phi_i$. The sum is over all matter fields. We normalize the generators such that for $\SU(N)$, the fundamental representation has $T_2(\square)= \frac1 2$ and $T_2(\adj) = N$.

A chiral primary operator $\cO$ with dimension $D[\cO]$ has R-charge
\be
R[\cO] = \frac23 \, D[\cO] = \frac23 \, \Big( D_{UV}[\cO] + \frac{\gamma[\cO]}2 \Big) \;.
\label{R.vs.gamma}
\ee
We can use this to recast the numerator of (\ref{nsvz}) as
\be
\label{ranom}
3T_2(\adj) - \sum_i T_2(\rep{r}_i) \big( 1 - \gamma_i(g_p) \big) = 3T_2(\adj) + 3 \sum_i R_i \, T_2(\rep{r} _i) = 3 \tr R \, T^a T^b \;,
\ee
where the trace is over the Weyl fermions in the theory. $T_2(\adj)$ is absorbed into this trace via the gauginos, which have unit R-charge. This result is just the statement that the superconformal R-anomaly is in the same multiplet as the trace of the stress tensor. Indeed, away from the fixed point, (\ref{ranom}) is still true just by supersymmetry. The only restriction is that the R-symmetry that shows up in the trace is the unique superconformal R-symmetry. This anomaly will not in general vanish away from the conformal point.

The situation in which we are interested here is where we have a Lagrangian theory whose gauge group is coupled to a non-Lagrangian sector.
In particular, consider a non-Lagrangian sector which has a flavor symmetry with
current superfield $\calJ^a$ that we couple to the gauge group via
\be
\label{GGM term}
\calL \,\supset\, 2 \int d^4\theta\,  \calJ^a \, \calV^a + \text{(terms for gauge invariance)} \;,
\ee
where ${\cal V}^a$ are the vector superfields of the Lagrangian sector.%
\footnote{Clearly the same formalism can be used if a Lagrangian \emph{is} available. For instance, for a set of chiral superfields $\Phi_i$, we can write $\; \calJ^a = \sum_i \Phi_i^\dag T^a \Phi_i$.}
It may be useful to think of this as the coupling between a visible (Lagrangian) sector and a hidden (non-Lagrangian) sector, as described in \cite{Meade:2008wd}.

The current multiplet $\calJ^a$ is a real linear superfield of dimension two satisfying
$ D^2 \calJ^a = \bar D^2 \calJ^a = 0$,
and containing the conserved current $j^a_\mu$ in the $\theta \sigma_\mu \bar\theta$ component.
When the flavor symmetry is a simple Lie group $G$, the OPE is that of a current algebra:
\be
\label{opeca}
j_\mu^a(x) \, j_\nu^b(0) = \frac{3k_G}{4\pi^4} \, \delta^{ab} \, \frac{x^2 g_{\mu\nu} - 2x_\mu x_\nu}{x^6} + \frac{2}{\pi^2} \, f^{abc} \, \frac{x_\mu x_\nu \, x \cdot J^c(0)}{x^6} + \dots
\ee
The coefficient $k_G$ is called
the central charge of the flavor symmetry.
In this normalization, $n$ free chiral multiplets have  $k_{\U(n)} = 1$,
and the contribution to $k_G$ for $G \subset U(n)$
depends on the embedding via
\begin{equation}
\label{kg}
k_G = 2 \sum T_2(\rep{r}_i) \;,
\end{equation}
where the fundamental decomposes as $\rep{n} = \sum_i \rep{r}_i$. If there is a flavor symmetry $H$ with a weakly gauged subgroup $G$, the central charge $k_{G \subset H}$ is given by
\be
k_{G \subset H} = I_{G \hookrightarrow H} \, k_H \;,
\ee
where $I_{G \hookrightarrow H}$ is the embedding index of $G$ in $H$, as defined in \cite{Argyres:2007cn}.%
\footnote{Given any representation \rep{r} of $H$ which decomposes into $\oplus_i \rep{r}_i$ of $G$, $ I_{G \hookrightarrow H} = {\sum_i T(\rep{r}_i)}/{T(\rep{r})}$.}
For more details on how to compute these quantities, see \cite{Argyres:2007cn,Barnes:2005bm}. Eq.~$(\ref{opeca})$ is important because the one-loop $\beta$-function corrects the gauge propagator. The contribution from the hidden sector is computed by the two-point function of the flavor current, and is proportional to $k_G$.

It is worth describing both the one-loop $\beta$-function as well as the exact one; the one-loop answer is useful \eg{} for determining when a theory is asymptotically free. In the following, we adopt holomorphically (as opposed to canonically) normalized gauge fields, so as to get rid of the denominator in the NSVZ formula \cite{ArkaniHamed:1997mj}.
The one-loop answer is obtained by setting $\gamma_i = 0$ in the numerator of the NSVZ $\beta$-function.
In the present case, we obtain
\be
\label{oneloop}
\beta_\text{one-loop} = 3 T_2(\adj) - \sum\nolimits_i T_2(\rep{r}_i) - \frac {k_G}2 \;,
\ee
where we have used (\ref{kg}) to write the sum over the hidden sector degrees of freedom in terms of its central charge (see App.~\ref{app: betafuncs} for an alternative derivation). This $\beta$-function is the only contribution to the running of the holomorphic gauge coupling, which is exact at one loop.

The physical gauge coupling receives contributions to all orders. To make things easy to keep track of, we will separate the contribution of the Lagrangian sector from that of the non-Lagrangian one. We define the 't Hooft anomaly coefficient $K$ from
the non-Lagrangian sector by the relation
\be
3 \tr\nolimits_\text{non-Lagrangian}\, R \, T^a T^b = -K \, \delta^{ab}. \label{K}
\ee
Then the exact NSVZ $\beta$-function is
\be
\label{exactnsvz}
\beta_{8 \pi^2 /g^2} = 3 \tr R \, T^a T^b = 3 T_2 (\adj) + 3 \sum_i R_i \, T_2(\rep{r}_i) - K \;,
\ee
where the second term comes from matter fields in the Lagrangian sector.

It might be instructive to use \eqref{R.vs.gamma} to rewrite \eqref{K} as \begin{equation}
-K \, \delta^{ab}= -\frac{k_G}{2}\delta^{ab} +\tr \gamma T^a T^b.
\end{equation} Then one can think of the first term on the right hand side as the one-loop contribution,
and of the second term as the contribution from the anomalous dimensions of the hidden sector.

\subsection{Examples}
\label{sec: beta examples}

Let us now describe some simple examples to illustrate the utility of (\ref{exactnsvz}). We will make use of the fact that the R-symmetry of a superconformal \Nugual{2} theory is $\SU(2)\times \U(1)$, and the superconformal \Nugual{1} $\U(1)_R$ inside it is
\be
R_{\cN=1} = \frac13 \, R_{\cN=2} + \frac43 \, I_3 \;,
\ee
where $I_3 = {\rm diag}(\frac12,-\frac12)$ is the generator of the Cartan subalgebra of the $\SU(2)$, and $R_{\cN=2}$ is the additional $\U(1)$. Note that any traces involving odd powers of $I_3$ will automatically vanish. See App.~\ref{app: formulae} for additional details.

\paragraph{\Nugual{2} theories.} When the theory has  \Nugual{2} supersymmetry, 
the superalgebra enforces
\be
\tr R_{\Nugual{2}} T^a T^b = - \frac{k_G}{2}\delta^{ab}
\ee
for any flavor symmetry $G$. From the expression for $R_{\Nugual{1}}$ we get $K = k_G/2$, and the exact $\beta$-function (\ref{exactnsvz}) stops at one-loop. 
There are no anomalous dimensions, because
the hypermultiplet is known not to be renormalized perturbatively.

\paragraph{Argyres-Seiberg theory.} In \cite{Argyres:2007cn} the authors considered an \Nugual{2} $\SU(2)$ gauge theory with one hypermultiplet, where $\SU(2)$ is a gauged subgroup of the $E_6$ flavor symmetry of the theory of MN.
Let us check that $\beta_{NSVZ} = 0$. In our normalization, $T_2(\SU(2)) = 2$, $T_2(\square) = 1/2$, the current algebra central charge $k_{E_6} = 6$, and the embedding index $I_{SU(2) \hookrightarrow E_6} = 1$. Applying (\ref{exactnsvz}) we get:
\be
\beta = 3 T_2(\adj) + 3 \sum\nolimits_i R_i \, T_2(\rep{r}_i) - \frac{k_{G}}2 = 3 \cdot 2 - 2- 1 - 3 = 0 \;.
\ee
The terms in the second to last equation are from the gaugino, adjoint fermion, hypermultiplet fermions and current algebra, respectively. All anomalous dimensions vanish.

Another example in \cite{Argyres:2007cn} features gauging an $\SU(2)$ subgroup of the $E_7$ flavor symmetry of the theory of MN, with no additional matter. In this case the central charge $k_{SU(2) \subset E_7} = 8$, so
\be
\beta = 3 \cdot 2 - 2 - 4 = 0 \;.
\ee

\paragraph{Mass deformed Argyres-Seiberg theory.} Consider the Argyres-Seiberg $E_6$ theory, mass deformed by $\delta W = m \Phi^2$ where $\Phi$ is the chiral superfield inside the vector multiplet. We integrate out $\Phi$ and as a result get a quartic superpotential for the hypermultiplet.
 In this case, the R-symmetry preserved by the mass deformation is (see \cite{Tachikawa:2009tt})
\be
\label{R symmetry mass deformation}
R_\mr{IR} = \frac12\, R_{\Nugual{2}} + I_3 = \frac32\, R_{\Nugual{1}} - I_3 \;,
\ee
where $R_{\Nugual{2}}$ and $R_{\Nugual{1}}$ refer to the UV \Nugual{2} theory, while $R_\mr{IR}$ refers to the IR \Nugual{1} theory. Then $K$, defined via $R_\mr{IR}$, is
\be
\label{kk}
K = \frac3 4\, k_G \;.
\ee
Then the $\beta$-function of $\SU(2)$ is
\be
\beta = 3 \cdot 2 - \frac32 - \frac34 \cdot 6 = 0 \;.
\ee

%
\section{Case study: $T_N$ theory coupled to $\cN=1$ vector multiplets}
\label{sec: TN}
%

In this section we study in detail a large class of \Nugual{1} superconformal
theories obtained by mass deforming the \Nugual{2} theories of Gaiotto \cite{Gaiotto:2009we}.
We identify their gravity duals and perform various checks, including a description of the manifold
of exactly marginal deformations as well as the
central charges $a$ and $c$.

%
\subsection{Rudiments of the $T_N$ theory}
\label{sec: TN review}
%

We begin by reviewing the basic properties of the $T_N$ theory \cite{Gaiotto:2009we, Gaiotto:2009gz}.
This is an $\cN=2$ superconformal field theory with no marginal couplings and
with flavor symmetry (at least) $\SU(N)^3$.
In particular $T_2$ just consists of eight free chiral multiplets $Q_{ijk}$,
while $T_3$ is the $E_6$ theory of Minahan-Nemeschansky.

The Coulomb branch of $T_N$ is parameterized by
dimension-$k$ operators $u_k^{(i)}$ for $k=3,4,\ldots,N$ and $i=1,2,\ldots,k-2$.
The structure of the Higgs branch is not fully understood, but it is known
that the chiral ring has dimension-two operators $\mu_{a}$  for $a=1,2,3$, each transforming in the adjoint
of the $a$-th $\SU(N)$ flavor symmetry. In addition, we have two dimension-$(N-1)$ operators
$Q_{ijk}$ and $\tilde Q^{ijk}$ transforming in the $(\mathbf{N},\mathbf{N},\mathbf{N})$
and $(\overline{\mathbf{N}},\overline{\mathbf{N}},\overline{\mathbf{N}})$ of the $\SU(N)^3$ symmetry.
It would be worthwhile to work out chiral ring relations among these operators, but for our purposes
we will only need the relation
\begin{equation}
\tr \mu_1^2=\tr \mu_2^2=\tr \mu_3^2 \;,
\end{equation}
whose derivation can be found in Appendix~\ref{app: chiral ring relations}.

It is useful for our purposes here to treat $T_N$ as an $\cN=1$ theory.
As explained in Appendix~\ref{app: formulae},
one linear combination of the $\cN=2$ $\U(1)_R$ and $\SU(2)_R$ symmetries
becomes a flavor symmetry $J$ from the $\cN=1$ point of view. The charges of various operators are:
\begin{equation}
\label{jcharge}
J(u_k^{(i)})=2k \;, \qquad\quad
J(\mu_i) = -2 \;, \qquad\quad
J(Q_{ijk})=J(\tilde Q^{ijk})=-(N-1) \;.
\end{equation}
The flavor symmetry central charge $k_G$ is $2N$ for any of the three $\SU(N)$ flavor symmetries.
As in the previous section, $k_G$ is related to the 't Hooft anomaly coefficient of the R-symmetry and
the $\SU(N)$ flavor symmetry via \eqref{k via N=1}.

\begin{figure}
\centering
\includegraphics[width=.6\textwidth]{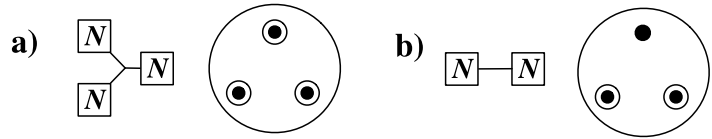}
\caption{a) The $T_N$ theory is shown as a triskelion on the left. The box with $N$ inside stands for an $\SU(N)$ flavor symmetry.
It is obtained by wrapping $N$ M5-branes on a sphere with three maximal punctures, as shown on the right.
b) A hypermultiplet in the bifundamental of $\SU(N)\times \SU(N)$
is shown as an edge with two boxes attached.
It is obtained by wrapping $N$ M5-branes on a sphere with two maximal punctures
and one simple puncture, denoted by $\bullet$.
\label{fig:blocks}}
\end{figure}

The $T_N$ theory is obtained by wrapping $N$ M5-branes on a sphere
with three maximal punctures, see Fig.~\ref{fig:blocks} a).
It is usually denoted by a triskelion, \ie{}~a trivalent vertex with three legs;
each box with $N$ in it signifies an $\SU(N)$ flavor symmetry. By comparison,
a bifundamental of $\SU(N)\times\SU(N)$ arises by wrapping the same number
of M5-branes on a sphere with two maximal punctures and one simple puncture,
see Fig.~\ref{fig:blocks} b). It is usually denoted by an edge having two boxes at the ends.
There, a maximal puncture is marked by $\odot$ and a simple puncture by $\bullet$;
one maximal puncture carries an $\SU(N)$ flavor symmetry,
and one simple puncture a $\U(1)$ symmetry.

\begin{figure}
\centering
\includegraphics[width=.6\textwidth]{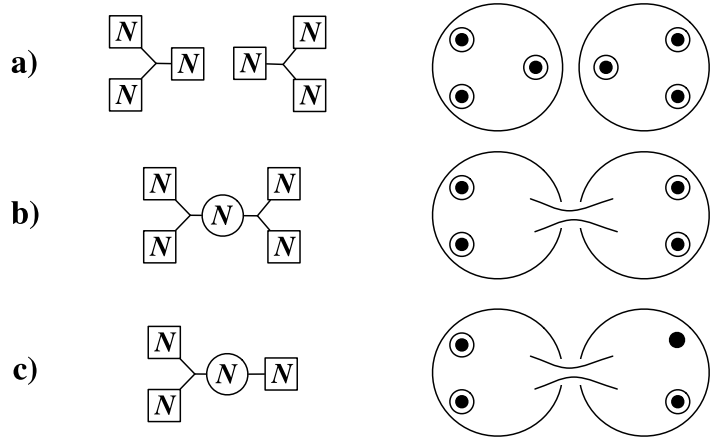}
\caption{a) Two copies of the $T_N$ theory. b) Two copies of $T_N$ with one
$\SU(N)$ gauge group, which couples to one from three $\SU(N)$
flavor symmetries for each of the $T_N$ theories. It arises from a sphere
with four maximal punctures.
c) One $T_N$ theory and a  bifundamental multiplet.
It arises from a sphere with three maximal punctures and one simple puncture.
\label{fig:connecting}}
\end{figure}

Now we can couple gauge fields to copies of $T_N$ and bifundamental fields,
as in  Fig.~\ref{fig:connecting}. An $\SU(N)$ gauge group is shown as a circle
with $N$ inside, and the fact that it couples to one of the $\SU(N)$ symmetries of
a $T_N$ theory is depicted by connecting it to one of the legs of a trivalent vertex
representing the $T_N$ theory, see Fig.~\ref{fig:connecting} b).
A weakly coupled $\SU(N)$ gauge group corresponds to a thin, long neck
developing on a Riemann surface as shown there.
Similarly, one can connect a bifundamental and a $T_N$ theory as shown in
Fig.~\ref{fig:connecting} c).

\begin{figure}[t]
\begin{center}
\includegraphics[height=.2\textheight]{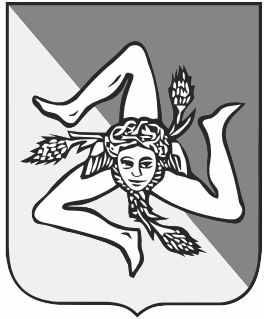}
\caption{The Sicilian flag. The central symbol is a triskelion. Image taken from Wikipedia.\label{fig: Sicilia_Stemma}}
\end{center}
\end{figure}

These are the simplest examples of the theories constructed by Gaiotto \cite{Gaiotto:2009we},
in which copies of $T_N$ theories and bifundamentals
are connected by $\SU(N)$ gauge groups coupling to the diagonal subgroup
of two $\SU(N)$ flavor symmetries. We dub these theories \emph{Sicilian gauge theories}
because their basic building block is a triskelion, the symbol of Sicily (see Figure \ref{fig: Sicilia_Stemma}). In contrast to Sicilian gauge theories, standard quiver theories
consist only of bifundamental fields and gauge groups.

%
\subsection{\Nugual{2} Sicilian gauge theories}
\label{sec: N=2 MN}
%

\begin{figure}
\centering
\includegraphics[width=.6\textwidth]{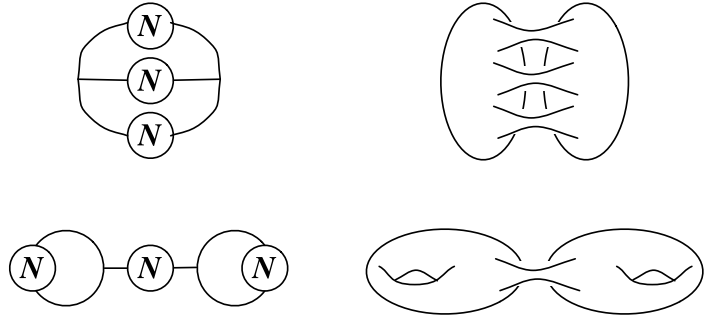}
\caption{The Sicilian gauge theory corresponding to $N$ M5-branes wrapped on a genus-$g$
Riemann surface $\Sigma_g$ without any punctures, for $g=2$. The left column
shows the Sicilian diagrams, and the right column shows the
corresponding degenerations of the Riemann surfaces.\label{fig:genus2}}
\end{figure}

Consider $2n$ copies of  a $T_N$ theory. This has $\SU(N)^{6n}$ flavor symmetry.
We can distribute these $6n$ $\SU(N)$ factors into $3n$ pairs,
and couple the diagonal subgroup of each pair $\SU(N)\times \SU(N)$ to
an $\cN=2$ gauge multiplet with gauge group $\SU(N)$.
The resulting theory has (generically) no flavor symmetries left.
For instance, all Sicilian diagrams for the case $n=1$
are shown in Fig.~\ref{fig:genus2}. As in the previous section, this corresponds to
$N$ M5-branes wrapped on a Riemann surface $\Sigma_g$ of genus $g=n+1$, but without punctures.
More precisely, the M5-branes wrap the zero section of the cotangent bundle
$T^*\Sigma_g$,  preserving $\cN=2$ supersymmetry. We will call such theories $\calT_g$.%
\footnote{The same notation will be used for \Nugual{1} examples, and we will explicitly specify the amount of supersymmetry.}

We can easily check that each of the gauge couplings is marginal, by using our results from Section \ref{sec: beta}.
These theories have no anomalous dimensions for the operators in the $T_N$ blocks,
and the exact NSVZ $\beta$-function vanishes:
\begin{equation}
\beta= 3T(\adj) - T(\adj) - 2 [ k_{\SU(N)}/2 ]  = 0 \;.
\end{equation}
The first two terms are the contributions from the \Nugual{2} vector multiplet,
while the last term comes from the flavor symmetry central charge of the $T_N$ theory.
Therefore, there are $3n=3g-3$  exactly marginal deformations preserving \Nugual{2} supersymmetry. This number nicely matches the moduli space dimension of the genus-$g$ Riemann surface.

The central charges of a general $T_N$ theory are \cite{Gaiotto:2009gz}
\begin{equation}
a_{T_N}=\frac{N^3}{6}-\frac{5N^2}{16}-\frac{N}{16}+\frac{5}{24} \;, \qquad\qquad
c_{T_N}=\frac{N^3}{6}-\frac{N^2}{4}-\frac{N}{12}+\frac{1}{6} \;.
\end{equation}
They are related to the 't Hooft anomalies of the R-symmetries
via the standard formul\ae{} \eqref{ac via N=2}. For the theories $\calT_g$ in question here,
the central charges can be easily calculated by adding the contributions from
copies of the $T_N$ theory and from the $\SU(N)$ vector multiplets. We find
\begin{equation}
\label{ac N=2 FT}
a= (g-1) \, \frac{8N^3 - 3N - 5}{24} \;, \qquad\qquad
c= (g-1) \, \frac{2N^3 -N -1}{6} \;.
\end{equation}

Maldacena and Gaiotto \cite{Gaiotto:2009gz} identified the holographic dual of this theory
as the gravity solution found ten years ago by Maldacena and Nu\~nez \cite{Maldacena:2000mw}.
One starts from $N$ M5-branes wrapped on a Riemann surface $\Sigma_g$
of genus $g=n+1$, embedded
supersymmetrically inside $T^*\Sigma_g$, and  takes the near horizon limit.
The central charges $a$ and $c$ have been calculated from the gravity side, reproducing the $\calO(N^3)$ and $\calO(N)$ terms written above.
On the gravity side, $\Sigma_g$ is endowed with a constant negative curvature metric,
and thus the moduli of the solution are the complex structure moduli of the
Riemann surface. This moduli space has complex dimension $3g-3$, which nicely
agrees with the field theoretical result.
As another check, Maldacena and Gaiotto identified
an M2-brane wrapped on $\Sigma_g$
as the gauge-invariant operator
\be
\label{prodQ}
\Theta = \prod_{a=1}^{2n} Q_{ijk}^{(a)} \;,
\ee
where $a$ runs over the copies of the $T_N$ theory
and gauge indices are appropriately contracted.
Obviously this has dimension $2n(N-1)$, which matches the mass of the
wrapped M2-brane.

%
\subsection{Deformation by adjoint mass: \Nugual{1} Sicilian theories}
\label{sec:massive}
%

The $\calT_g$ theory above, in a duality frame where there are only $T_N$ blocks,
has an $\cN=1$ superpotential schematically of the form:
\begin{equation}
W = \sum\nolimits_s \tr  ( \Phi_s \, \mu_{a(s),i(s)}) -\tr (\Phi_s \, \mu_{b(s),j(s)}) \;,
\end{equation}
where $\Phi_s$ ($s=1,\ldots,3n$)
is the adjoint scalar in the $s$-th $\SU(N)$ vector multiplet
and $\mu_{a,i}$ ($a=1,\ldots,2n$, $i=1,2,3$) is the chiral operator
in the adjoint of the $i$-th $\SU(N)$ flavor symmetry of the $a$-th copy
of the $T_N$ theory. $a(s),i(s)$ and $b(s),j(s)$ specify which copy of the $T_N$ theory
the $s$-th $\SU(N)$ vector couples to.

Now let us add mass terms for the adjoint chiral multiplets $\Phi_s$:
\begin{equation}
\label{mass-deformation}
W_m = \sum\nolimits_s m_s^2 \tr \Phi_s^2 \;.
\end{equation}
At energy scales far below $m_s$, we can integrate $\Phi_s$ out. The superpotential becomes
\begin{equation}
\sum\nolimits_s \frac{1}{m_s} \, \tr \big( \mu_{a(s),i(s)}-\mu_{b(s),j(s)} \big)^2 \;.
\end{equation}
This suggests that the theory reaches a superconformal point where the operators $\mu_{a,i}$  all have dimension $3/2$.

We can indeed check that the NSVZ $\beta$-functions vanish. The anomalous dimension $\gamma$
can be identified with a multiple of the current $J$ in (\ref{jcharge}). In particular,
$\gamma= J /2,$ because the charge of $\mu$ under $J$ is $-2$.
The contribution to the beta function from the anomalous dimension is
\begin{equation}
\tr ( \gamma T^a T^b ) = \frac12 \tr (J T^a T^b) = - \frac{k_{\SU(N)}}4 \delta^{ab} \;.
\end{equation}
We can now evaluate the beta function:
\begin{equation}
\beta = 3T(\adj)-2[k_{\SU(N)}/2] - 2[k_{\SU(N)}/4] = 0 \;,
\end{equation}
where the first, second  and the third term
come from the \Nugual{1} vector multiplet, the one-loop contribution of
two $T_N$ theories coupled to the gauge group, and
the contribution from the anomalous dimensions of two $T_N$ theories, respectively.

Another way to state this result is to consider the low-energy R-symmetry
\be
R_\mr{IR} = \frac12\, R_{\Nugual{2}} + I_3 = R_\mr{UV} + \frac16 \, J \;,
\ee
found in \cite{Tachikawa:2009tt} as the combination of \Nugual{2} R-charges preserved by the mass
deformation (\ref{mass-deformation}). We use it in the formula of the exact NSVZ $\beta$-function
(\ref{exactnsvz}). Since $R_\mr{IR}$ is anomaly free with respect to all gauge groups,
\begin{equation}
\tr R_{IR} T^a T^b = 0 \;,
\end{equation}
so the $\beta$-function vanishes. This is similar to the last example in Section \ref{sec: beta examples}.

The central charges can now be evaluated from the standard formula \eqref{ac via N=1 R} (see also \cite{Tachikawa:2009tt}),
yielding
\begin{equation}
\label{ac N=1 FT}
a = (g-1) \, \frac{9N^3 - 3N - 6}{32} \;, \qquad\qquad
c = (g-1) \, \frac{9N^3 - 5N - 4}{32} \;.
\end{equation}
Finally, we note that the operator \eqref{prodQ} now has dimension
\begin{equation}
\label{dim operator theta N=1}
D[\Theta] = \frac32 \, n(N-1) \;.
\end{equation}

%
\subsection{Counting exactly marginal deformations}\label{sec:counting}
%

Naively, the $\calT_g$ theory after the $\cN=1$ deformation would have
$6n-1$ exactly marginal couplings: $3n$ come from the coupling constants
of the original $\cN=2$ theory, and $3n-1$ come from the ratio of the mass parameters.
We will see below that we have, in fact, one more
exactly marginal direction, and the total number is
\be
\label{number marginal deformations}
\dim_\bbC \calM_C = 6n \;.
\ee

In the following we will discuss the case of $N > 3$. The case $N=2$, worked out in \cite{Maruyoshi:2009uk}, needs a different discussion even though the result is substantially the same. The case $N=3$ has some peculiarities as well,
because the operators $Q_{ijk}$ and $\mu_i$ have the same dimension.

We perform an analysis as in Leigh and Strassler \cite{Leigh:1995ep}.
The candidate marginal deformations are
\bea
& \Tr (\mu_{a,i})^2 & & 2n \\
& \Tr \mu_{a(s),i(s)} \, \mu_{b(s),j(s)} \qquad & & 3n \\
& W_{(s)}^\alpha W_{(s),\alpha} &  & 3n \;,
\eea
where in the first column are the operators
and in the second column are their multiplicities.
The first are squares of the moment maps, three for each $T_N$ theory;
as explained in Appendix \ref{app: chiral ring relations},
the three squares are in fact the same in the chiral ring.
The second are products of moment maps on each gauge node; the third are gauge couplings.

To have zero NSVZ beta functions for all gauge groups,
the anomalous dimension of each $T_N$ theory is constrained to be $\gamma=J/2$.
This imposes $2n$ real conditions on the couplings.
At the same time, we can perform the flavor rotation of each $T_N$ theory
by the symmetry $J$. This will eliminate $2n$ phases.
Therefore  we end up with $6n$ exactly marginal parameters.%
\footnote{%
We would like to give a somewhat lengthy comment on the usage in the literature
of the method by Leigh and Strassler. We often find the line of arguments as follows.
Schematically, one starts by counting chiral operators
of dimension three which can be used to deform the superpotential;
let us say we have $n$ of them, including the complexified gauge couplings.
One then writes down the beta functions for the gauge and superpotential couplings.
Typically they boil down to a number, say $m$, of constraints on the anomalous dimensions
of chiral superfields. Then one concludes that there are $n-m$ exactly marginal directions,
because one needs to impose $m$ constraints which are functions of $n$ parameters.

This procedure is, however, not correct as stated.
The number $n$ is complex,  because the parameters
are either coefficients in the superpotential or complexified gauge couplings.
On the other hand, the anomalous dimensions are real functions and only give $m$ real constraints.
Therefore at this stage one can only say that there are $2n-m$ real degrees of freedom.
This might not be an even number, whereas we expect the exactly marginal
deformations to form a complex manifold which should have an even number of real dimensions.
In the examples treated in the original paper \cite{Leigh:1995ep}, one could perform
phase redefinitions of chiral superfields
to eliminate some phases from the superpotential couplings, possibly shifting the theta angles.
However, this redefinition cannot be done in a general field theory, which may have a complicated
combination of gauge groups and matter fields.

The point is that we need to identify theories related by field redefinitions,
typically by  anomalous or broken global symmetries,
which change the theta angles and the phases of superpotential couplings.
It seems to us that there are always $m$ such $\U(1)$ rotations when there are
$m$ constraints on the anomalous dimensions. At least, this is the case for the examples
we will deal with in our paper.
Then, we need to subtract $2m$ real degrees from $n$ complex degrees of freedom,
resulting in $n-m$ exactly marginal directions.

It would be interesting to demonstrate that the number of anomalous/broken $\U(1)$ rotations
and the number of constraints on the anomalous dimensions are always the same.
Furthermore, this procedure of imposing $m$ real conditions and then removing $m$ phases
is reminiscent of the K\"ahler quotient construction. It might be worthwhile to pursue this analogy further. For further discussion of such matters, see \cite{Aharony:2002hx, Kol:2002zt, Benvenuti:2005wi}.}

For $N=3$, the trifundamental operators $Q_{ijk}$ and $\tilde Q^{ijk}$ of dimension $(N-1)$ in the $T_3$ theory have dimension 2 as the moment maps $\mu_i$. In fact together they form the moment map of $E_6$. After an \Nugual{1} gauging they acquire anomalous dimension as $\mu_i$, allowing us in principle to construct one more candidate marginal operator for each $T_3$ block: $Q_{ijk} \tilde Q^{ijk}$. However, as shown in App.~\ref{app: chiral ring relations}, in the chiral ring of $T_3$ $Q_{ijk} \tilde Q^{ijk}$ is proportional to $\tr \mu_i^2$. The previous discussion is thus unchanged.%
\footnote{The only possible exception is $\calT_2$ and $N=3$, where we can contract $Q_{ijk}$ and $\tilde Q^{ijk}$ from two different $T_3$ theories.}

%
\subsection{Holographic duals}
%

As briefly reviewed in Sec.~\ref{sec: N=2 MN}, the holographic dual of the Sicilian \Nugual{2} $\calT_g$ theory (made of $2n$ copies of $T_N$ theory
glued together by $3n$ $\cN=2$  vector multiplets of $\SU(N)$ gauge groups)
was identified as the $\cN=2$ solution of 11d supergravity
found by Maldacena and Nu\~nez \cite{Maldacena:2000mw}.
Our aim here is to demonstrate that the mass-deformed $\cN=1$ version of the $\calT_g$ theory
has as its holographic dual the analogous $\cN=1$ supergravity solution, also found in \cite{Maldacena:2000mw}.

\paragraph{Construction of the gravity solution.}
Let us quickly recall
how the gravity solutions were constructed.
For a given Riemann surface $\Sigma_g$ of genus $g>1$, one endows $\Sigma_g$ with a metric
of constant negative curvature, and goes on to find a solution of the form $AdS_5\times \Sigma_g$
of 7d $\cN=4$ $\SO(5)$ gauged supergravity.
To preserve some supersymmetry, one embeds the spin connection of $\Sigma_g$
into the $\SO(5)$ R-symmetry gauge fields. We will briefly review this procedure below. More details can be found in Appendix \ref{app: twisting}.

In order to preserve $\cN=2$ supersymmetry in the dual SCFT,
one embeds the spin connection in the $\SO(2)$ part of
\begin{equation}
\label{decomposition2}
\SO(2)\times \SO(3) \;\subset\; \SO(5) \;.
\end{equation}
The spinors transform in the  \rep{4} of $\SO(5)$, and decompose as
\begin{equation}
\rep{4} \to \rep{2}_+ \oplus \rep{2}_-
\end{equation}
under \eqref{decomposition2}. Let us say $\rep{2}_+$ becomes
covariantly constant, thanks to the embedding of the spin connection.
We then find that
$\SO(2)\times\SO(3)$, which commutes with the $\SO(2)$ subgroup we just chose,
gives the $\U(1)\times\SU(2)$ R-symmetry of the $\cN=2$ SCFT.

Similarly, to have $\cN=1$ SUSY in the dual SCFT,
we decompose $\SO(5)$ as
\begin{equation}
\label{decomposition1}
\SU(2) \times\SU(2)_F \,\simeq\, \SO(4) \;\subset\; \SO(5)
\end{equation}
and embed the spin connection in $\U(1)_R \subset \SU(2)$.
The \rep{4} of $\SO(5)$ decomposes under $\U(1)_R \times \SU(2)_F$ as
\begin{equation}
\rep{4} \to \rep{1}_+ \oplus \rep{1}_- \oplus \rep{2}_0 \;,
\end{equation}
following \eqref{decomposition1}.
The twisting makes, say, $\rep{1}_+$ covariantly constant.
Then $\U(1)_R \subset \SU(2)$ gives the $\U(1)$
R-symmetry of the $\cN=1$ SCFT.

Notice that the $\SU(2)_F$ part remains unbroken.
The unbroken supersymmetry parameter is neutral under it,
which means that $\SU(2)_F$ becomes a non-R flavor symmetry of the dual SCFT. We will shortly see how
additional \Nugual{1} preserving parameters can break it.
For this class of solutions it is known how to lift
them to solutions of eleven-dimensional supergravity by adjoining a properly warped $\tilde S^4$ to the seven-dimensional spacetime
\cite{Cvetic:1999xp,Liu:1999ai}.
For completeness, the detailed form of the resulting metric is given in Appendix \ref{app: MN metric}.

\paragraph{$\SU(2)_F$ Wilson lines.}
The metric just mentioned (see App.~\ref{app: MN metric}) describes the product of $AdS_5$ with a non-compact
hyperbolic plane $H^2$, over which is fibred a squashed $\tilde S^4$; the isometry is $\U(1)_R \times \SU(2)_F$. To have a compact genus-$g$ Riemann surface,
we need to quotient by a discrete (Fuchsian) subgroup $\Gamma$ of $\SL(2,\bR)$.
This introduces $3g-3$ complex structure moduli.
These are the sole moduli in the case of the $\cN=2$ solution.
For the $\cN=1$ solution, we can introduce Wilson lines of the $\SU(2)_F$ field
discussed above without breaking the supersymmetry,
because the unbroken supersymmetry parameter is neutral under $\SU(2)_F$.
Of course this will generically break $\SU(2)_F$ completely, and in fact the dual field theory does not generically have any $\SU(2)_F$ flavor symmetry.
More explicitly, the supergravity solution is constructed by choosing a discrete
subgroup $\Gamma_W$ of $\SL(2,\bbR) \times \SU(2)_F$, with the requirement that the projection $\Gamma_W \to \SL(2,\bbR)$ is injective, and quotienting the fibre bundle $\tilde S^4 \to E \to H^2$ by it.

The number of degrees of freedom in the Wilson lines can be counted as follows:
The homotopy of $\Sigma_g$ is generated by
the loops $A_{1,\ldots,g}$ and $B_{1,\ldots,g}$ with one relation,
\begin{equation}
(A_1B_1A_1^{-1}B_1^{-1})(A_2B_2A_2^{-1}B_2^{-1})\cdots
(A_gB_gA_g^{-1}B_g^{-1}) = \unit \;.
\end{equation}
The Wilson lines are a homeomorphism from the homotopy group to $\SU(2)_F$.
There are three real degrees of freedom for each $A_i$ and $B_i$, with three removed by the relation above,
another three removed by the global $\SU(2)_F$ transformations. In total, we have $3\times 2g -3-3 = 6g-6$.
It is known that there is a natural complex structure
on the space of Wilson lines. Therefore, we have a complex parameter space of dimension
\be
\dim_\bbC \calM_\text{Wilson lines} = 3g-3 \;.
\ee
The number $n$, which specifies the number of copies of $T_N$ and of gauge groups, is related
to the genus by $g = n+1$. Thus, summing the $3g-3$ complex parameters of the Riemann surface and
the $3g-3$ parameters of the Wilson lines, we reproduce the $6n$ exactly marginal deformations
in the field theory (\ref{number marginal deformations}).

Even though we do not have a precise map from the space of gauge and superpotential couplings to
the space of Riemann surfaces with $\SU(2)_F$ Wilson lines, this construction shows that the S-duality
group should act on the former as it does on the latter.

\paragraph{Universality of the construction.}
The above construction gives, with regard to the complex structure, the most general
configuration with $N$ coincident M5-branes wrapped on a surface
preserving $\cN=1$ supersymmetry, as we now show.
\footnote{The authors thank David R. Morrison for helpful discussions on this point.}
Supersymmetry requires that the total space of the normal bundle of the M5-branes
is locally Calabi-Yau.
The fiber of the normal bundle is $\bC^2$,
and the curvature is in $\U(2)$. The Calabi-Yau condition
constrains the $\U(1)$ part of the curvature to be equal to that of the
canonical bundle, but the $\SU(2)$ part is unconstrained.
Therefore, the moduli of the system are given by the the moduli of the Riemann
surface plus the moduli of the $\SU(2)$ bundle over the surface. This is exactly what we found in our
supergravity construction.
This argument strongly suggests that, although we only considered a rather
restricted set of superpotential deformations in Sec.~\ref{sec:massive} and \ref{sec:counting},
the manifold of IR superconformal theories we found describes the entire
moduli space of M5-branes wrapped on a Riemann surface.

\paragraph{$\U(1)_F$ enhanced flavor symmetry.} The introduction of $\SU(2)_F$ Wilson lines
generically breaks such isometry completely. However we preserve (part of) it for particular choices
of the Wilson lines. A subgroup $\U(1)_F$ is preserved by commuting Wilson lines, whose parameter
space has complex dimension $g$; the full $\SU(2)_F$ is preserved for vanishing Wilson lines.
We expect the same flavor symmetry enhancement to take place in the field theory on particular
submanifolds of the conformal manifold $\calM_C$. Let us show the $\U(1)_F$ enhancement.

\begin{figure}[tn]
\begin{center}
\includegraphics[height=4.3cm]{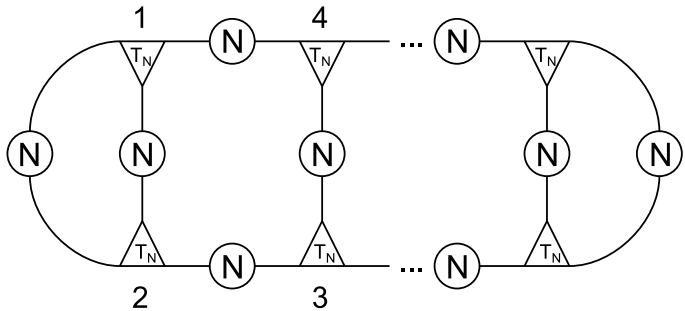}
\caption{The Sicilian $\calT_g$ theory in a particular S-dual frame. In this particular picture, $T_N$ blocks are emphasized as triangles, and numbered. The combination $Y = J^{(1)} - J^{(2)} + J^{(3)} - J^{(4)} + \dots$, where each $J^{(a)}$ term is the $\U(1)$ flavor symmetry of a $T_N$ theory, is the anomaly free $\U(1)_F$. \label{fig:U(1) enhancement}}
\end{center}
\end{figure}

Consider the \Nugual{1} $\calT_g$ theory in the particular S-dual frame of Figure \ref{fig:U(1) enhancement}. Before writing down any superpotential, there exists a unique anomaly-free $\U(1)_F$ flavor symmetry:
\be
Y = J^{(1)} - J^{(2)} + J^{(3)} - J^{(4)} + \dots
\ee
where the structure of the sum is understood in relation to Fig. \ref{fig:U(1) enhancement}. Each $J^{(a)}$ factor is the $\U(1)$ flavor symmetry of a copy of the $T_N$ theory, defined in (\ref{jcharge}) or (\ref{jcharge formula}).
We now look for the conformal manifold compatible with this $\U(1)_F$ symmetry. The candidate marginal deformations are:
\bea
& \Tr \mu_{a(s),i(s)} \, \mu_{b(s),j(s)} \qquad & & 3n \\
& W^\alpha_{(s)} W_{(s),\alpha} & & 3n \;.
\eea
The first are products of moment maps, one for each gauge group; the second are gauge couplings. The number of independent field redefinitions is now only $2n-1$, because there are $2n$ $J^{(a)}$ actions but one combination is an anomaly-free symmetry which does not affect either the superpotential nor the theta angles; the number of real relations on the anomalous dimensions is $2n - 1$ as well. We then find a conformal manifold of complex dimension
\be
\dim_\bbC \calM_C = 4n+1 = 3(g-1) + g \;.
\ee
This agrees with the moduli of a genus $g$ Riemann surface with commuting $\SU(2)_F$ Wilson lines.

We expect a further enhancement of the flavor symmetry to $\SU(2)_F$ on a smaller submanifold of dimension $3n = 3(g-1)$. We have not been able to show the existence of such a symmetry, which is likely realized only quantum mechanically at the fixed point.%
\footnote{Along the same lines, notice that there are different S-dual frames than that of Fig. \ref{fig:U(1) enhancement} where there is no manifest anomaly-free $\U(1)_F$, even before writing down any superpotential.}

\paragraph{M2-brane wrapped on $\Sigma_g$.}
Let us next compare the dimension of the operator $\Theta$ in \eqref{prodQ}, which is dual to an
M2-brane wrapped on $\Sigma_g$.
On the gravity side, the dimension is proportional to the product
of the radius of AdS$_5$ and the area of the Riemann surface. Fixing the coefficient,
we find
\begin{equation}
D = 8 (g-1) X_0^{1/2} e^{f+2h} N\;,
\end{equation}
whose constants are defined in App.~\ref{app: MN metric}.
Then we conclude:
\begin{equation}
D_{\cN=2}=2(g-1)N + \calO(1) \;, \qquad\qquad
D_{\cN=1}=\frac32\, (g-1)N + \calO(1) \;.
\end{equation}
The order 1 (in $N$) corrections are expected from quantization of zero modes on the M2-brane, as
in \cite{Gaiotto:2009gz}.
The holographic dual correctly reproduces
the dimension of this operator as below (\ref{prodQ}) and in (\ref{dim operator theta N=1}), to leading order in $N$.

\paragraph{Central charges $a$ and $c$.}
We can also match the central charges. At leading order in $N$,
the gravity side predicts $c = {\pi R_{\mr{AdS}_5}^3}/{8 G_N^{(5)}}$ \cite{Henningson:1998gx} (where $G_N^{(5)}$ is the five dimensional Newton constant) which results in \cite{Maldacena:2000mw}
\begin{equation}
a=c=\frac83 \, (g-1) \, e^{3f+2h} N^3 \;.
\end{equation}
See App.~\ref{app: MN metric} for the constants entering into the metric. We then get:
\begin{equation}
\label{ac leading contribution}
a_{\cN=2} = c_{\cN=2}=\frac13 \, (g-1) N^3 \;, \qquad\qquad
a_{\cN=1} = c_{\cN=1} = \frac9{32} \, (g-1) N^3 \;.
\end{equation}
The subleading terms can be found from the wrapped M5-branes.
We will focus on the anomalies $\tr R_{\Nugual{1}}$ and $\tr R_{\Nugual{1}}^3$,
which are related to $a$ and $c$ by (\ref{ac via N=1 R}).

First let us recall the relation between the anomaly coefficients
and the anomaly polynomial, for a 4d field theory (for a review of the analysis
of the anomaly, see \eg{} \cite{Harvey:2005it}).
The anomaly polynomial for one Weyl fermion of charge $q$ is
\begin{equation}
I_6 = \ch (qF) \, \hat A(T_4) \bigm|_6 = \frac {q^3}6 \, c_1(F)^3  - \frac {q}{24} \, c_1(F) \, p_1(T_4) \;,
\end{equation}
where $F$ stands for the $\U(1)$ bundle of the $\U(1)$ charge, and $T_4$ for the tangent bundle of spacetime. $\ch$ is the Chern character, $\hat A$ the Dirac genus, $c_k$ the Chern classes and $p_k$ the Pontryagin classes.
Therefore, a theory with R-anomaly $\tr R^3$ and $\tr R$ has the anomaly polynomial:
\begin{equation}
I_6= \frac{\tr R^3 }{6} \, c_1(F)^3 -\frac{\tr R}{24} \, c_1(F) \, p_1(T_4) \;.
\end{equation}
The anomaly eight-form for $N$ M5-branes was found in \cite{Witten:1996hc,Harvey:1998bx}, from certain Chern-Simons terms in 7d supergravity. It is:
\begin{equation}
\label{HMM anomaly polynomial}
I_8(N) = N \, I_8(1) + (N^3-N) \, \frac{p_2(N)}{24} \;,
\end{equation}
where
\begin{equation}
I_8(1) = \frac{1}{48} \left[ p_2(N) - p_2(T) + \frac14 \big( p_1(T) - p_1(N) \big)^2 \right]
\end{equation}
is the anomaly for a single M5-brane.
Here $N$ and $T$ stand for the normal and tangent bundle of the worldvolume of the M5-branes, respectively. Let us also recall that $p_{1,2}(B)$ are given, in terms of the skew Chern roots $\pm b_i$
of the bundle $B$, as
\begin{equation}
p_1(B) = \sum_i b_i^2 \;, \qquad\qquad p_2(B) = \sum_{i<j} b_i^2 b_j^2 \;.
\end{equation}

Now let us reduce the anomaly eight-form on the genus-$g$ Riemann surface.
We denote the Chern roots of the tangent bundle as $\pm \lambda_1$, $\pm \lambda_2$,
$\pm t$ and those of the normal bundle as $\pm n_1$, $\pm n_2$;
here $\lambda_i$ are along the 4d spacetime while $t,n_{1,2}$ are along the Riemann surface.
We denote by $F$ the $\U(1)$ bundle which couples to the R-symmetry;
as discussed in Appendix \ref{app: twisting}, this enters by shifting $n_{1,2}$ as
\begin{equation}
n_1 \to n_1 + c_1(F) \;, \qquad\qquad n_2 \to n_2 + c_1(F) \;.
\end{equation}
\Nugual{1} supersymmetry requires
\begin{equation}
n_1 + n_2 + t = 0
\end{equation}
and the fact that $\Sigma_g$ is genus $g$ implies
\begin{equation}
\int _{\Sigma_g} t = 2-2g \;.
\end{equation}
The relevant term in the anomaly polynomial is linear in $t$ and, after integrating over $\Sigma_g$, we get:
\begin{equation}
\int_{\Sigma_g} I_8 = \frac16 \, (g-1) \, N^3 \, c_1(F)^3 - \frac{1}{24} \, (g-1) \, N \, c_1(F) \, p_1(T_4) \;.
\end{equation}
Therefore we have:
\begin{equation}
\tr R_{\Nugual{1}}^3 = (g-1) \, N^3 \;, \qquad\qquad \tr R_{\Nugual{1}} = (g-1) \, N \;.
\end{equation}
Using \eqref{ac via N=1 R} we find that this nicely reproduces the field theory central charges $a$, $c$ \eqref{ac N=1 FT}, up to $\cO(1)$ corrections.

Indeed the anomaly polynomial (\ref{HMM anomaly polynomial}) refers to the 6d theory on the worldvolume of $N$ M5-branes, comprised of the non-trivial SCFT dual to the near-horizon geometry plus the decoupled free dynamics of the center of mass. Led by this consideration, we observe that if we remove the anomaly of a single M5-brane (whose low-energy dynamics is free), that is if we start from
\be
I_8[A_{N-1}] = I_8(N) - I_8(1) \;,
\ee
we exactly reproduce the field theory result (\ref{ac N=1 FT}).%
\footnote{For completeness, let us observe that the $\Nugual{2}$ charges (\ref{ac N=2 FT}) can be exactly reproduced in a similar way. Starting from $I_8[A_{N-1}]$, we include the $\U(1)$ bundle of the $R_{\Nugual{1}}$ charge via $n_1 \to n_1 + \frac23\, c_1(F)$, $n_2 \to n_2 + \frac43\, c_1(F)$. \Nugual{2} SUSY requires $n_1 + t = 0$ and $n_2 = 0$. This leads to $\tr R = \frac23 (N-1)(g-1)$, $\tr R^3 = \frac{2}{27} (16N^3 - 3N - 13)(g-1)$, which gives (\ref{ac N=2 FT}). The bundle of $R_{\Nugual{2}}$ can be analyzed with $n_1 \to n_1 + 2c_1(F)$.}

%
\subsection{Theories with maximal punctures}
%

Let us now study the Sicilian \Nugual{1} SCFTs obtained from $N \geq 3$ M5-branes wrapped on $\Sigma_g$ with $n_N$ punctures of maximal type. We will dub them $\calT_{g,n_N}$. Again, the case $N=2$ has been considered in \cite{Maruyoshi:2009uk}, while $N=3$ does not exhibit any further subtleties with respect to the previous section. We require $n_N \geq 3$ if $g=0$, or $n_N \geq 1$ if $g=1$ to make the situation generic.

The field theory is constructed from $2n + n_N$ copies of $T_N$, by gauging together $3n + n_N$ pairs of $\SU(N)$ global symmetries (we consider connected Sicilian diagrams). The resulting $\calT_{g,n_N}$ theory has then (at least) $\SU(N)^{n_N}$ flavor symmetry, and the genus is $g=n+1$. The candidate marginal deformations are:
\bea
& \Tr (\mu_{a,i})^2 & & 2n + n_N \\
& \Tr \mu_{a(s),i(s)} \, \mu_{b(s),j(s)} \qquad & & 3n + n_N \\
& W_{(s)}^\alpha W_{(s),\alpha} & & 3n + n_N\;.
\eea
As before, the first are squares of the moment maps, three for each $T_N$ theory but equal in the chiral ring; the second are products of moment maps on each gauge node; the third are gauge couplings. These deformations preserve the $\SU(N)^{n_N}$ flavor symmetry. From $\gamma(\mu_{a,i}^m) = -1$ we get $2n + n_N$ real relations and as many axial rotations. Thus we see that the number of exactly marginal deformations is
\be
\dim_\bbC \calM_C = 6n + 2n_N = 6(g-1) + 2n_N \;.
\ee

In the \Nugual{2} case \cite{Gaiotto:2009gz}, the gravity dual to the $\calT_{g,n_N}$ theory is easily constructed. We enrich the Fuchsian group $\Gamma \subset \SL(2,\bbR)$, by which we quotient $H^2$ in order to obtain a compact Riemann surface, in such a way that the fundamental domain has $n_N$ $\bbZ_N$ orbifold singularities. Due to the \Nugual{2} twist, the identification $z \simeq e^{2\pi i/N} z$ (where $z$ is a local coordinate on $H^2$) is accompanied by a rotation in the fibre and the singularity is locally $\bbC^2/\bbZ_N$.

This is consistent, because along an $A_{N-1}$ singularity in M-theory live $\SU(N)$ gauge fields, which realize an $\SU(N)$ global symmetry factor in the boundary theory. Such singularity can arise as the uplift of type IIA D6-branes. We can understand it as follows. In the IIA construction of \cite{Witten:1997sc} a global $\SU(N)$ comes from $N$ D4-branes stretching between one NS5-brane and $N$ D6's; uplifting to M-theory we get the near-horizon geometry of $N$ M5's in the presence of an $A_{N-1}$ singularity. More generally, in the IIB construction of \cite{Benini:2009gi} a generic puncture is realized by $N$ $(p,q)$ 5-branes partitioned and ending on some number of 7-branes. Dualizing to M-theory, they map to uplifted D6's, whose number and structure matches with the results in \cite{Gaiotto:2009gz}.

In the \Nugual{1} case, the singularity corresponding to a maximal puncture must look locally the same, \ie{} like $\bbC^2/\bbZ_N$. The group element needs to act as
\begin{equation}
(z,\, x_1,\, x_2) \,\to\, (e^{2\pi i/N} z,\,  e^{-2\pi i/N} x_1,\, x_2) \;,
\end{equation}
where $z$ is a complex coordinate on the hyperbolic plane
centered around the orbifold point, and $x_1$, $x_2$ are appropriate coordinates
on the $\bC^2$ fiber. The \Nugual{1} twist \eqref{decomposition1} alone would give
the action
\begin{equation}
(z, \, x_1,\, x_2) \,\to\, (e^{2\pi i/N} z\, , e^{-\pi i/N} x_1,\, e^{-\pi i/N} x_2) \;.
\end{equation}
This means that the maximal puncture requires an $\SU(2)$ monodromy around it, with
fixed conjugacy class
\begin{equation}
(x_1,\, x_2) \,\to\, (e^{-\pi i/N} x_1, \, e^{\pi i/N} x_2) \;.
\end{equation}
Such monodromy is an $\SU(2)_F$ Wilson line around the puncture, and the choice of a particular element in the conjugacy class is a free complex parameter. The supergravity solution is obtained from the metric in App.~\ref{app: MN metric}, as in the previous section, by quotienting the bundle $\tilde S^4 \to E \to H^2$ by a subgroup $\Gamma_W$ of $\SL(2,\bbR) \times \SU(2)_F$ with the properties stated above.

The moduli of the solution are the following: $3(g-1) + n_N$ complex structure moduli of a genus-$g$ Riemann surface with $n_N$ indistinguishable punctures, plus $3(g-1) + n_N$ Wilson lines on it. This matches the field theory result. With maximal punctures the $\SU(2)_F$ Wilson lines cannot be turned off, and the maximal symmetry enhancement is $\U(1)_F$, which happens on a $3(g-1) + n_N + g$ dimensional submanifold of $\calM_C$. It is easy to check, at least in some S-duality frames, that the same happens in the field theory.

Notice that \Nugual{1} supersymmetry allows more general orbifold singularities (\eg{} $\bbC^3/\bbZ_k \times \bbZ_l$) with respect to the \Nugual{2} case. It would be interesting to understand their dual field theory.

%
\subsection{Theories with simple punctures}
%

The \Nugual{1} theory $\calT_{g,n_1}$ on $N \geq 3$ M5-branes wrapped on $\Sigma_g$ with $n_1$ simple $\U(1)$ punctures, has associated Sicilian diagrams of different types. The ones with all $\SU(N)$ gauge groups are made by gauging together $2n$ $T_N$ theories and $n_1$ bifundamental blocks with $\U(1)$ flavor symmetry (as in Fig. \ref{fig:blocks} b), through $3n + n_1$ $\SU(N)$ gauge blocks. The flavor symmetry is $\U(1)^{n_1}$ and the genus $g=n+1$.

Among the $\SU(N)$ gauge blocks, it is better to distinguish if they join two bifundamentals, two $T_N$ blocks, or one for each type. We introduce the number $b$:
\begin{center}
\begin{tabular}{cccc}
\tabs two bifundamentals & $b$ & \multirow{2}{*}{$\Big\} \; 2n_1 - b$} & \multirow{3}{*}[-.5ex]{$\Bigg\} \; 3n + n_1$} \\
\tabs $T_N$ and bifundamental & $2(n_1 - b)$ & & \\
\tabs two $T_N$ & \multicolumn{2}{c}{$3n + (b-n_1)$} &
\end{tabular}
\end{center}
The number $b$ is not invariant under S-duality and is subject to some constraints, however
the final result does not depend on it so we will not be concerned about this. We have the following candidate marginal deformations:
\bea
& \Tr (\mu_{a,i})^2 & & 2n \\
& \Tr \mu_{a(s),i(s)} \, \mu_{b(s),j(s)} & & 3n + (b-n_1) \\
& \Tr \mu_{a(s),i(s)} \, Q^{s,t} \, \tilde Q^{s,t} & & 2(n_1 - b) \\
& \Tr ( Q^{s,t} \, \tilde Q^{s,t} )^2 & & n_1 \\
& \Tr Q^{s,t} \, \tilde Q^{s,t} \, Q^{t,u} \, \tilde Q^{t,u} & & b \\
& \big( \Tr Q^{s,t} \, \tilde Q^{s,t} \big) \, \big( \Tr Q^{u,v} \, \tilde Q^{u,v} \big) \qquad & & n_1(n_1+1)/2 \\
& W_{(s)}^\alpha W_{(s),\alpha}  & & 3n + n_1 \;.
\eea
We label bifundamentals as $Q^{s,t}$, where the indices refer to the gauge groups joined by them.
The novelty here is that there are double trace deformations.

From $\gamma(\mu_{a,i}) =-1$, $ \gamma(Q^{s,t}) = -1/2$  we get $2n + n_1$ real relations.
We can also perform the axial rotation by the symmetry $J$ for each copy of $T_N$,
and rotate the bifundamentals, eliminating $2n + n_1$ phases in total.
We conclude that the number of exactly marginal deformations is:
\be
\label{conformal dim simple punctures}
\dim_\bbC \calM_C = 6(g-1) + 2n_1 + \frac{n_1(n_1 +1)}{2} \;.
\ee

The geometric interpretation is as follows.
The moduli of the Riemann surface $\Sigma_g$ with $\SU(2)_F$ Wilson lines provide $6(g-1)$ parameters.
Each $\U(1)$ puncture provides two complex parameters:\footnote{The authors thank J. Maldacena for insights on this point.}
Since the $\U(1)$ puncture corresponds to the intersection
of an extra M5-brane with the stack of $N$ M5-branes,
one parameter is for its position on the Riemann surface $\Sigma_g$,
the other for the $\mathbb{CP}^1$ worth of directions of the M5-brane inside the two-dimensional fiber over $\Sigma_g$.
If we take the near-horizon limit of the
$N$ M5-branes, the transverse M5's can be dealt with in the probe brane approximation.
Their worldvolume fills AdS$_5$ at a specified point on the Riemann surface,
and wraps a circle on $\tilde S^4$ parameterized by $\mathbb{CP}^1$.

Finally, there are  $n_1 (n_1 + 1)/2$  ``double-trace'' exactly marginal deformations
constructed from $n_1$ gauge-invariant chiral primary operators of dimension $3/2$.
In the holographic dual, these deformations correspond to the mixed boundary conditions
imposed on the bulk fields dual to these $n_1$ operators \cite{Witten:2001ua}. It would be nice
to understand what are they in the 6d $(2,0)$ theory.

\

\begin{figure}
\centering
\includegraphics[width=.6\textwidth]{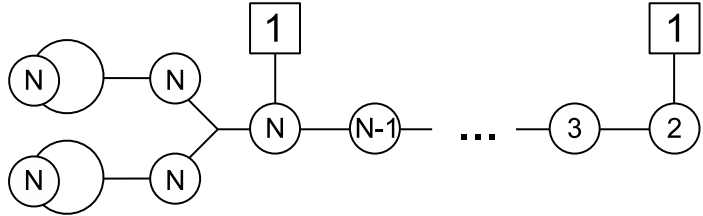}
\caption{$\calT_{g=2,\, n_1 = N}$ theory, in an S-duality frame which exhibits a tail.
\label{fig: Tailg=2}}
\end{figure}

With simple punctures different Sicilian diagrams are possible, in which some gauge ranks are smaller than $N$. In particular there can be superconformal tails. An example for $\calT_{g=2,\, n_1=N}$ is in Figure \ref{fig: Tailg=2}, and some more examples are in \cite{Gaiotto:2009we, Gaiotto:2009gz}. The reader can fill the details of the computation, whose result is still (\ref{conformal dim simple punctures}).

Along the same lines, one can consider $\calT_{g,(n_N,n_1)}$, arising from a Riemann surface with $n_N$ maximal and $n_1$ simple punctures. Both the supergravity and field theory computation yield $\dim_\bbC \calM_C = 6(g-1) + 2n_N + 2n_1 + n_1(n_1+1)/2$.

%
\section{Conclusions}
\label{sec: conclusion}
%
In this work, we have provided a general set of tools which can be used to analyze supersymmetric field theories in which non-Lagrangian sectors are coupled to gauge groups. Although we have only analyzed a particular subset of such theories here, our results are general, and could be used to analyze (and, indeed, construct) many new theories. It is no doubt an interesting pursuit to try to find some new and exotic theories, and investigate their superconformal phases via the above techniques.

As an example of a large new set of interesting SCFTs, we considered in particular the $\cN=1$ theories resulting from mass deformations of the $\cN=2$ theories pioneered by Gaiotto, and also described their holographic duals. We then provided a number of checks on the duality, by computing central charges as well as the dimension of the conformal manifold. It is interesting that gravity sides of some of these holographic duals were constructed many years ago by Maldacena and Nu\~{n}ez, although the corresponding SCFTs remained mysterious until the present work.

There are many possible directions for future study. As mentioned above, it is particularly interesting to see what kinds of new Sicilian SCFTs we can construct. Although finding the gravity duals of these theories is in general no doubt a difficult problem, the techniques presented here at least make it possible to perform checks of the correspondence. On the other hand, our supergravity analysis revealed that intrinsically \Nugual{1} punctures are possible, and finding their field theory dual is an intriguing problem. Another task is to clarify the shape of the M5-branes wrapping the Riemann surface on the whole moduli and parameter space, whose equations are controlled by VEVs in the $(2,0)$ theory. Finally it could be interesting to see if the reduction to 3d, as done in \cite{Gaiotto:2009hg} for the \Nugual{2} case, holds any new surprises.

\section*{Acknowledgments}
The authors thank Y. Ookouchi for collaboration at the early stage of this work.
The authors have greatly benefited from discussions with S. Benvenuti, D. Gaiotto, N. Halmagyi, J. Maldacena and M. Strassler.
FB and YT acknowledge the kind hospitality of the Aspen Center for Physics, and FB that of the Benasque Science Center, during which this work was pursued.

FB is supported by the NSF grant PHY-0756966.
YT is supported in part by the NSF grant PHY-0503584, and by the Marvin L.
Goldberger membership at the Institute for Advanced Study.
BW is supported in part by DOE grant DE-FG02-90ER40542, and by the Frank and Peggy Taplin Membership at the Institute for Advanced Study.

\appendix

\section{Standard formul\ae{} for SCFTs}
\label{app: formulae}

\subsection{$\cN=1$ SCFTs}

The central charges $a$ and $c$
are defined as coefficients of terms in the conformal anomaly of the trace of the energy-momentum tensor
generated by a background gravitational field:
\begin{equation}
\vev{T_\mu^\mu} = \frac{c}{16\pi^2} \, (\text{Weyl})^2 - \frac{a}{16 \pi^2} \, (\text{Euler}) \;,
\end{equation}
where
\be
({\rm Weyl})^2 = R^2_{\mu\nu\rho\sigma} - 2R^2_{\mu\nu} + \frac13 R^2 \;, \qquad\qquad
({\rm Euler})= R^2_{\mu\nu\rho\sigma}-4R^2_{\mu\nu}+ R^2 \;.
\ee
For $\cN=1$ SCFTs, they are related to the
't Hooft anomalies of the $\U(1)$ R-symmetry in the superconformal algebra
via the relations \cite{Anselmi:1996dd, Anselmi:1997am}
\begin{equation}
\label{ac via N=1 R}
a = \frac{3}{32} \, \big[ 3\tr R_{\cN=1}^3-\tr R_{\cN=1} \big] \;, \qquad\qquad
c = \frac{1}{32} \, \big[ 9\tr R_{\cN=1}^3 - 5\tr R_{\cN=1} \big] \;.
\end{equation}
We use the convention that
$\tr J_1J_2J_3 $ stands for the 't Hooft anomaly coefficient among three currents
of the theory, normalized so that it will be equal to the trace over the label of Weyl fermions
when the theory has a weakly-coupled Lagrangian description.

For a flavor symmetry $G$, the central charge $k_G$ is defined as the coefficient of the
two-point function of currents of $G$ as in \eqref{opeca}.
It is related to the 't Hooft anomaly via
\begin{equation}
\label{k via N=1}
k_G \, \delta^{AB} = -6 \tr (R_{\cN=1} T^A T^B)
\end{equation}
where $T^A$ are the generators of the flavor symmetry $G$. We normalize them
so that $T^A$ in the fundamental of $\SU(2)$ have eigenvalues $\pm1/2$.

A chiral primary operator $\cO$ has dimension
\begin{equation}
D[\cO] = \frac32 \, R_{\cN=1} [\cO] \;.
\end{equation}

\subsection{$\cN=2$ SCFTs}
The R-symmetry of $\cN=2$ SCFTs is $\U(1)_R\times \SU(2)_R$.
We denote the generators as $R_{\cN=2}$ and $I^a$,  where $a=1,2,3$.
Three-point functions of the R-currents and the energy-momentum tensor
are known to contain only two superconformal invariants \cite{Kuzenko:1999pi}.
One consequence is the relation of the central charges $a$ and $c$ to the anomaly:
\begin{equation}
\label{ac via N=2}
\tr R_{\cN=2}^3 = \tr R_{\cN=2} = 48(a-c) \;, \qquad\qquad
\tr R_{\cN=2} I^a I^b = \delta_{ab} (4a-2c) \;.
\end{equation}
Similarly, we have the relation of the flavor central charge and the anomaly
\begin{equation}
\label{k via N=2}
k_G \, \delta^{AB}= -2 \tr (R_{\cN=2} T^A T^B) \;.
\end{equation}

A chiral primary operator $\cO$ on the Coulomb branch has dimension
\begin{equation}
D[\cO] = \frac12 \, R_{\cN=2}[\cO] \;,
\end{equation}
and carries no spin under $\SU(2)_R$. On the other hand,
a chiral primary operator on the Higgs branch has $R_{\cN=2} = 0$
and the dimension is twice the $\SU(2)_R$ spin.

\begin{table}
\[
\begin{array}{c|ccc}
\tabs R_{\cN=2} \; \backslash \; I^3 & \tfrac12 & 0 & -\tfrac12 \\
\hline
0 & & A_\mu \\
1 & \lambda_\alpha & & \lambda'_\alpha \\
2 & & \phi
\end{array} \qquad\qquad
\begin{array}{c|ccc}
\tabs R_{\cN=2} \; \backslash \; I^3 & \tfrac12 & 0 & -\tfrac12 \\
\hline
-1 & & \psi_\alpha \\
0 & Q & & \tilde Q^\dag \\
1 & & \bar{\tilde\psi}_{\dot\alpha}
\end{array}
\]
\caption{Charges of $\cN=2$ free multiplets.
Note that $A_\mu$ and $\lambda_\alpha$ form an $\cN=1$ vector multiplet,
while $\phi$ and $\lambda'_\alpha$ form an $\cN=1$ chiral multiplet.\label{table: free charges}}
\end{table}

It is sometimes useful to treat a given $\cN=2$  theory as an $\cN=1$ SCFT,
by fixing a particular $\cN=1$ subalgebra inside the $\cN=2$ superconformal algebra.
The $\cN=1$ $\U(1)_R$ symmetry is given by the linear combination
\begin{equation}
\label{R N=1 combination}
R_{\cN=1} = \frac13 \, R_{\cN=2} + \frac43 \, I_3 \;,
\end{equation}
while another linear combination
\begin{equation}
\label{jcharge formula}
J = R_{\cN=2} - 2I_3
\end{equation}
commutes with the chosen $\cN=1$ subalgebra,
and is thus a flavor symmetry from the $\cN=1$ point of view.
One can easily check the statement
using the well-known assignment of charges for free hypermultiplets and vector multiplets
in Table~\ref{table: free charges}.

\section{Derivation of the generalized NSVZ $\beta$-function}
\label{app: betafuncs}

We give here a more detailed alternative derivation of the exact $\beta$-function formula (\ref{exactnsvz}) which generalizes the NSVZ expression to non-Lagrangian sectors. We adopt the approach of Arkani-Hamed and Murayama \cite{ArkaniHamed:1997mj}. We start by reviewing their derivation to fix conventions, and then show how to include the contribution of superconformal non-Lagrangian sectors.

\subsection{Theories with Lagrangian descriptions}

Consider a Lagrangian theory consisting of vector multiplets $\calV^a$ of a gauge group $G$,
chiral multiplets $\Phi_i$ transforming in a representation $\rep{r}_i$ of $G$
and a (possibly zero) superpotential.
We can think of this system as a ``matter theory'' (the fields $\Phi_i$) whose flavor symmetry group $H$ is (partially)
gauged by $\calV^a$ according to $G \subset H$, as in (\ref{GGM term}). In this case the current superfield coupled to $\calV^a$ is simply $\calJ^a = \sum_i \Phi_a^\dag T^a \Phi_i$, $T^a$ being the generators of $G$.

We consider a Wilsonian action defined at some cut-off $M$. We choose to holomorphically normalize the gauge fields. The Lagrangian then contains the holomorphic gauge coupling $g_h$ (as opposed to the physical gauge coupling $g_p$). We compute the variation in the couplings as we vary the cut-off to $M'$ while keeping the IR physics fixed.
Take the Lagrangian at cut-off $M$ to be
\be
\label{Lagrangian gauging}
\calL_h^M =\int d^4\theta\,  \Phi_i^\dag \, e^{2\calV_h^i} \, \Phi_i   +  \left [ \frac1{16} \int d^2\theta\, \frac{1}{g_h^2} \, W^a W^a + \text{h.c.}  \right ] + \dots
\ee
Here the subscript $h$ stands for ``holomorphically normalized.''
The dots stand for the remaining Lagrangian including superpotential, non-canonical K\"ahler potential, \etc.~, which we abbreviate since they do not play any role in the discussion. The sum over $i$ is implicit.

In this normalization, the perturbative running of the gauge coupling is exact at one-loop.
The running of the complexified gauge coupling
\begin{equation}
\frac{1}{g_h^2} = \frac{1}{g^2} + i \, \frac{\theta}{8\pi^2}
\end{equation}
is necessarily holomorphic, which is only true when $\parfrac{}{\log M} \, \frac{8\pi^2}{g_h^2} = b_0$ is a constant (because it must be independent of the $\theta$-angle). This implies that the renormalized Lagrangian at cut-off $M'$ is
\be
\calL_h^{M'} = \int\!\! d^4\theta\, Z_i(M,M') \, \Phi_i^\dag e^{2\calV_h^i} \Phi_i + \left [ \frac1{16} \int\!\! d^2\theta\, \Big( \frac{1}{g_h^2} - \frac{b_0}{8\pi^2} \log \frac{M}{M'} \Big) \, W^a W^a + \text{h.c.} \right ] + \dots
\ee
where $b_0 = 3T_2(\adj) - \sum_i T_2(\rep{r}_i)$ is the one-loop coefficient of the $\beta$-function. The change in the Wilsonian action can only be holomorphic if we allow the coefficient of the matter kinetic terms (which are manifestly non-holomorphic) to change from 1 to $Z_i(M,M')$. This is just the wave-function renormalization.

To study the running of the physical gauge coupling, we need to keep vector multiplets and matter fields
canonically normalized. The change of normalization of the vector multiplets, due to an anomalous Jacobian, produces the denominator of the NSVZ formula \cite{ArkaniHamed:1997mj}. However, we prefer to keep the vector multiplets holomorphically normalized and only insist on having the matter fields canonically normalized.%
\footnote{This is customary in the literature. One of the reasons is that this is the normalization usually obtained from supergravity, and thus the corresponding $\beta$-function is the one usually read holographically.}
Therefore we perform the change of variables $\Phi_i = Z_i^{-1/2} \Phi_i'$, which yields an anomalous Jacobian in the path integral.
The Jacobian can be computed by noticing that we can write $Z = e^{i\alpha}$: for real $\alpha$ the contribution to the path integral is the usual chiral anomaly. The F-term part of the result for real $Z$ is then obtained by holomorphy. For real $\alpha$ the exact result is
\be
\calD( e^{-\frac{i\alpha}2} \Phi') \calD( e^{\frac{i\alpha}{2}} \bar\Phi') = \calD\Phi' \calD\bar\Phi' \, \exp\Big\{ \frac1{16} \int\!\! d^4y\, d^2\theta \, \frac{T_2(\rep{r}_\Phi)}{8\pi^2} \, \log(e^{-i\alpha}) \, W^aW^a + \text{h.c.} \Big\} \;.
\ee
For complex $\alpha$ there are D-terms generated which cannot be derived simply by analytic continuation. However, these terms do not affect the $\beta$-function. The F-terms can be inferred by holomorphy.
The final result is
\be
\calL_p^{M'} = \int d^4\theta\,  \Phi_i^\dag \, e^{2\calV_h^i} \, \Phi_i + \left [ \frac1{16} \int d^2\theta\, \frac{1}{g_p'^2} \, W^a W^a + \text{h.c.} \right ] + \dots \;,
\ee
with
\be
\frac{1}{g_p'^2} = \frac{1}{g_h^2} - \frac{b_0}{8\pi^2} \, \log\frac{M}{M'} - \sum_i \frac{T_2(\rep{r}_i)}{8\pi^2} \, \log Z_i(M,M') \;.
\ee
Here $g_p$ stands for the physical coupling, even though with a slight abuse of terminology because we keep the vector multiplets holomorphically normalized.
Using the definition of anomalous dimension $\gamma_i \equiv \partial\log Z_i/ \partial\log M$, we get
\be
\label{holomorphic NSVZ}
\beta_{8\pi^2/g^2} \equiv \parfrac{}{\log M} \, \frac{8\pi^2}{g_p^2} = 3T_2(\adj) - \sum_i T_2(\rep{r}_i) \big( 1 - \gamma_i \big) \;.
\ee
This expression is the numerator of the NSVZ $\beta$-function.

\subsection{Theories with non-Lagrangian sector}

We now generalize the above argument to the case of a non-Lagrangian theory with some flavor symmetry $H$, of which we gauge a subgroup $G \subset H$. We will proceed as if a Lagrangian exists but it is not known. As in \cite{Meade:2008wd}, the gauging corresponds to coupling the vector multiplets $\calV^a$ with the current superfields $\calJ^a$ of the flavor symmetry $G$:
\be
\calL \supset 2 \int d^4\theta\,  \calJ^a \, \calV^a + \text{(terms for gauge invariance)} \;.
\ee
The OPE of the current superfield in (\ref{opeca}) is controlled by $k_G$, the central charge of the current algebra. \Nugual{1} supersymmetry relates the central charge and the 't Hooft anomaly involving the $\cN=1$
R-symmetry through
\begin{equation}
k_G \, \delta^{ab} = -6 \tr R_{\cN=1} T^a T^b \;.
\end{equation}
The trace on the right hand side is over the Weyl fermions in the theory \emph{if} it has a Lagrangian description;
otherwise it needs to be defined abstractly as the coefficient of the three-point function of currents.
It will be important that these traces can be calculated
even for non-Lagrangian theories, thanks to the robustness of the anomaly coefficients
against quantum corrections.

As before, we change the cut-off of the Wilsonian Lagrangian from $M$ to $M'$ keeping the IR physics fixed, and look at the behavior of the gauge coupling. First we keep both the gauge multiplet and the hidden sector fields {\em holomorphically normalized}.
Then the change in $1/g_h^2$ is exhausted at one-loop in $g_h$. The one-loop contribution of the hidden sector can be computed, as in \cite{Meade:2008wd, Argyres:2007cn}, from two-point functions of the current multiplet.
We have
\be
\frac{1}{g_h'{}^2} =\frac{1}{g_h^2} - \frac{b_0}{8\pi^2} \, \log \frac{M}{M'} \;,
\ee
where $b_0 = 3T_2(\adj) - k_G/2 $ and $k_G$ is the flavor central charge of
the hidden sector.

We then need to take into account the fact that a chiral  operator $\cO$ in the hidden sector
will receive a wave-function renormalization. In the standard case, we have \begin{equation}
\Phi_i \to \Phi_i'=Z_i(M,M')^{1/2}\, \Phi_i.
\end{equation} Generalizing this, let us define the wave-function renormalization factor
of $\cO$ as  
\begin{equation}
\label{baz}
\cO' = Z_{\cO}(M,M')^{1/2} \, \cO\;.
\end{equation}
This is how an operator $\cO$ gets the anomalous dimension
$D_\mr{IR}[\cO]= D_\mr{UV}[\cO]+\gamma[\cO]/2$  with
\begin{equation}
\gamma[\cO] = \frac{\partial \log Z_{\cO}}{\partial\log M} \;.
\end{equation}
In general there is nothing like canonical normalization for a non-Lagrangian sector, because there are no preferred operators. However, when the sector is conformal, we know that there exists a rescaling that keeps all operators fixed.
Suppose the transformation
\begin{equation}
\cO \;\to\; \exp(i\epsilon\gamma[\cO]) \, \cO
\end{equation}
for real $\epsilon$ is a global symmetry $\gamma$ of the hidden sector.
This transformation shifts the $\theta$ term of the gauge group by the amount proportional to the 't Hooft anomaly $ \tr \gamma T^a T^b$.
By holomorphy for imaginary $\epsilon$, $1/g_p^2$ gets a shift proportional to the anomaly.
It is straightforward to fix the coefficient, and we find:
\begin{align}
\label{strongly coupled NSVZ}
\parfrac{}{\log M} \, \frac{8\pi^2}{g_p^2} \, \delta^{ab} &= b_0 \, \delta^{ab} +  \tr \gamma T^a T^b \nonumber \\
&= 3T(\adj)\, \delta^{ab} + 3 \tr R_{UV} T^a T^b + \tr \gamma T^a T^b \;.
\end{align}
When the theory in the infrared is superconformal, the last two terms combine into $3\tr R_{IR} T^a T^b$
because $3R_{IR} = 2 D_{IR} = 2 (D_{UV}+\gamma/2) = 3R_{UV} + \gamma$.
Then the $\beta$-function is proportional to the 't~Hooft anomaly of the R-symmetry, and its vanishing is consistent with the existence of an anomaly-free R-symmetry.

Alternatively, suppose the superconformal hidden sector has a Lagrangian description but it is not known. Then we can directly take the expression (\ref{holomorphic NSVZ}) and recast it in the form
\be
\beta_{8\pi^2/g^2} = 3T_2(\adj) - K
\ee
where $K$ is defined as $3\Tr R\, T^a T^b = -K \, \delta^{ab}$. $K$, being the coefficient of the 't Hooft anomaly, is defined and computable independently of the Lagrangian. Since it does not matter if the Lagrangian is known or exists at all, the expression above is what we are looking for, and is formula (\ref{exactnsvz}).
Let us conclude by noticing that such an expression requires the non-Lagrangian sector to be superconformal (otherwise there is no concept of canonical normalization), but the gauge coupling $g_p$ can run.

\section{A chiral ring relation of $T_N$}
\label{app: chiral ring relations}

The theory $T_N$ has $\SU(N)^3$ flavor symmetry;
correspondingly, its chiral ring (as an $\cN=1$ theory)
has chiral operators $\mu_{1,2,3}$ of dimension two,
transforming in the adjoint of each of the three $\SU(N)$ symmetries.
We argue that $\tr \mu_1^2=\tr \mu_2^2 = \tr\mu_3^2$
at the level of chiral ring relations.

\subsection{$N=2$}
The theory $T_2$ consists of eight $\cN=1$ chiral multiplets
 $Q_{ijk}$, $i,j,k=1,2$; three $\SU(2)$ act on $i,j,k$ respectively.
 Then $\mu_{1,ii'}=Q_{ijk} Q_{i'j'k'} \epsilon^{jj'}\epsilon^{kk'}$, and similarly for $\mu_2$ and $\mu_3$.
It is easy to see that $\tr \mu_1^2 = \tr \mu_2^2=\tr \mu_3^2$.
In fact there is only one quartic singlet operator constructed out of $Q_{ijk}$.

\subsection{$N>2$}
$T_N$ with $N>2$ itself is hard to analyze. Instead
let us couple it to a superconformal tail
\begin{equation}
\label{aho}
\SU(N-1)\times \SU(N-2)\times \cdots \times \SU(2) \;.
\end{equation}
Then it is S-dual to the standard linear quiver
\begin{equation}
\SU(N)_1\times \cdots \times \SU(N)_{N-2} \;.
\end{equation}
We have bifundamental fields
$Q_{\alpha}$ and  $\tilde Q_\alpha$
for $\alpha=1,2,\ldots,N-1$, where
$Q_{\alpha}$  and $\tilde Q_\alpha$  are charged under
$\SU(N)_{\alpha-1}$ and $\SU(N)_{\alpha}$.
Here $\SU(N)_0$ and $\SU(N)_{N-1}$ are the two $\SU(N)$ flavor
symmetries.
Let $V_\alpha$ be the fundamental representation of $\SU(N)_\alpha$.
Then we  regard $Q_{\alpha}$ and $\tilde Q_{\alpha}$ as linear maps
\begin{equation}
Q_\alpha:  V_{\alpha-1} \to V_{\alpha} \;, \qquad\qquad
\tilde Q_\alpha:  V_{\alpha} \to V_{\alpha-1} \;.
\end{equation}
We identify
\begin{equation}
\mu_1 = \tilde Q_1 Q_1 - \frac1N \tr \tilde Q_1 Q_1 \;, \qquad\qquad
\mu_2 = Q_{N-1} \tilde Q_{N-1} - \frac1N \tr  Q_{N-1} \tilde Q_{N-1} \;.
\end{equation}

We would like to show that
\begin{equation}
\label{zot}
\tr \mu_1^2=\tr \mu_2^2 \;,
\end{equation}
which is
\begin{multline}
\tr \tilde Q_1 Q_1\tilde Q_1 Q_1 - \frac1N (\tr \tilde Q_1Q_1)^2\\
=\tr Q_{N-1} \tilde Q_{N-1} Q_{N-1}\tilde Q_{N-1} - \frac1N (\tr Q_{N-1}\tilde Q_{N-1})^2 \;.
\end{multline}
This follows if we can show
\begin{multline}
\tr \tilde Q_\alpha Q_\alpha\tilde Q_\alpha Q_\alpha
 - \frac1N (\tr \tilde Q_\alpha Q_\alpha)^2\\
=\tr Q_{\alpha-1} \tilde Q_{\alpha-1} Q_{\alpha-1}\tilde Q_{\alpha-1} - \frac1N (\tr Q_{\alpha-1}\tilde Q_{\alpha-1})^2\label{foo}
\end{multline}
for $\alpha=1,\ldots,N-2$. This last equality follows from the F-term
relation for the adjoint scalar $\Phi_\alpha$
of the $\SU(N)_\alpha$ gauge multiplet.
Indeed, the superpotential is
\begin{equation}
\tr \tilde Q_\alpha  \Phi_\alpha Q_\alpha + \tr  Q_{\alpha+1}  \Phi_\alpha \tilde Q_{\alpha+1} \;.
\end{equation}
Recalling that $\Phi_\alpha$ is traceless, the F-term relation is
\begin{equation}
Q_{\alpha}\tilde Q_\alpha -\frac1N\tr Q_{\alpha}\tilde Q_\alpha = - \Big[
\tilde Q_{\alpha+1}Q_{\alpha+1} -\frac1N\tr \tilde Q_{\alpha+1} Q_{\alpha+1} \Big] \;.
\end{equation}
Taking the trace of the square of both sides,
we obtain \eqref{foo}, proving \eqref{zot}.
We believe that this chiral ring relation comes from the chiral ring of $T_N$ itself.

\subsection{$N=3$}

For $N=3$ we can explicitly prove the relation.
The theory $T_3$ is the $E_6$ theory of Minahan-Nemechansky,
whose chiral ring has been studied in \cite{Gaiotto:2008nz}. We can repeat the argument
using the subgroup $\SU(3)^3\subset E_6$.
The chiral ring, on the hyperk\"ahler side, is generated by
a set of dimension-two operators $\bX$ transforming in the adjoint of $E_6$
which satisfy the quadratic relations
\begin{equation}
\label{bar}
(\bX \otimes  \bX)|_{\cI_2} = 0 \;,
\end{equation}
where the projection is on the representation $\cI_2$ defined by the relation
\begin{equation}
\Sym^2 V(\adj) = V(2\,\adj) \oplus \cI_2 \;.
\end{equation}
Here $V(\alpha)$ is the representation with highest weight $\alpha$, and $\adj$ stands for the highest weight of the adjoint representation. Thus
$V(\adj) = \rep{78}$ is the adjoint representation of $E_6$ whose Dynkin label is
\begin{equation}
\adj = \Esix010000 \;.
\end{equation}
We then have
\begin{equation}
\cI_2 =  V\left(\Esix100001\right)\oplus V\left(\Esix000000\right) = \rep{650} \oplus \rep{1} \;.
\end{equation}

Under the $\SU(3)^3$ subgroup, $\bX$ decomposes into
$\mu_{1,2,3}$, each adjoint of one $\SU(3)$, and $Q_{ijk}$ and $\tilde Q^{ijk}$:
\begin{equation}
\rep{78} = (\rep{8},\rep{1},\rep{1}) \oplus (\rep{1},\rep{8},\rep1) \oplus (\rep1,\rep1,\rep{8}) \oplus (\rep{3},\rep{3},\rep{3}) \oplus (\bar {\rep{3}},\bar {\rep{3}},\bar {\rep{3}}) \;.
\end{equation}
The product $(\bX\otimes  \bX)$
contains four $\SU(3)^3$ singlets,
\ie{} $\tr\mu_1^2$, $\tr\mu_2^2$, $\tr\mu_3^2$ and $Q_{ijk}\tilde Q^{ijk}$.
On the other hand decomposing $\cI_2$ we find three singlets, which means that there are three linearly-independent
$\SU(3)^3$-invariant relations. We conclude that
the four operators just listed are all proportional.

\section{Twisting the $(2,0)$ theory}
\label{app: twisting}

The 6d $(2,0)$ theory has $\OSp(6,2|4)$ superconformal invariance, whose bosonic subgroup is the 6d conformal group times $\USp(4) \simeq \SO(5)$ R-symmetry. In particular, there are supercharges $Q$ and real scalar fields $\Delta$ transforming under $\SO(5,1) \times \SO(5)_R$ as:
\be
Q: \; \rep{4} \otimes \rep{4} \quad \text{(with symplectic Majorana condition)} \;, \qquad\qquad \Delta: \; \rep{1} \otimes \rep{5} \;.
\ee
There is also a two-form potential $B_{MN}$ whose field strength is self-dual, singlet of $\SO(5)_R$, and spinors transforming as $\rep{4'} \otimes \rep{4}$.

We put the theory on a Riemann surface $\Sigma_g$, and twist it embedding the spin connection $\SO(2)_s$ of $\Sigma_g$ into $\SO(5)_R$.

\subsection{\Nugual{2} twist} We embed $\SO(2)_s$ into the $\SO(2)_R$ factor of $\SO(2)_R \times \SO(3)_R \subset \SO(5)_R$. Before the twisting the supercharges transform under $\SO(3,1) \times \SO(2)_s \times \SO(3)_R \times \SO(2)_R$ as
\be
\rep{4} \otimes \rep{4} \quad\to\quad \big( \rep{2}_\frac{1}{2} + \rep{2}'_{-\frac1 2} \big) \otimes \big( \rep{2}_\frac{1}{2} + \rep{2}_{- \frac{1}{2}} \big) \;,
\ee
and the symplectic Majorana condition reduces to a relation between \rep{2} and \rep{2'} of $\SO(3,1)$. We twist the spin connection as $\SO(2)_s \to \SO(2)_s - \SO(2)_R$. The supercharges become
\be
\rep{2}_0 \otimes \rep{2}_{\frac1 2} + \rep{2}_1 \otimes \rep{2}_{-\frac1 2} + \rep{2}'_{-1} \otimes \rep{2}_{\frac1 2} + \rep{2}'_0 \otimes \rep{2}_{-\frac1 2} \;.
\ee
The preserved supercharges (covariantly constant  on the Riemann surface) are $\rep{2}_0 \otimes \rep{2}_{\frac1 2}$, which generate an \Nugual{2} superalgebra. $R_{\Nugual{2}}$ has to be identified with twice the charge under $\SO(2)_R$, so that it is correctly normalized as $R[Q]=1$. $\rep{2}'_0 \otimes \rep{2}_{-\frac1 2}$ are just the conjugate $Q^\dag$. The scalars $\Delta$ decompose as:
\be
\rep{1} \otimes \rep{5} \quad\to\quad \rep{1}_0 \otimes \big( \rep{3}_0 + \rep{1}_1 + \rep{1}_{-1} \big) \quad\xrightarrow{\text{twisting}}\quad \rep{1}_0 \otimes \rep{3}_0 + \big( \rep{1}_{-1} \otimes \rep{1}_1 \big)_\bbC \;.
\ee
$(\rep{1}_{-1} \otimes \rep{1}_1)_\bbC$ is a complex scalar field; it is a holomorphic differential on the Riemann surface, with $R = 2$ and dimension 1. $\rep{1}_0 \otimes \rep{3}_0$ is an $\SU(2)_R$ triplet of real scalar fields, scalars on the Riemann surface, with $R = 0$. They pair up with the $\SU(2)_R$ singlet $B_{ab}$ (two legs on the Riemann surface) and with $\rep{2}_0 \otimes \rep{2}_{-\frac12}$, an $\SU(2)_R$ doublet of fermions with $R=-1$, to form a tensor multiplet. The twisting of the fermions $\rep{4}' \otimes \rep{4}$ is worked out easily, giving $\rep{2}_{-1} \otimes \rep{2}_{\frac12} + \rep{2}_0 \otimes \rep{2}_{-\frac12}$ and their conjugates.


\subsection{\Nugual{1} twist}
We now embed $\SO(2)_s$ into the $\U(1)_R$ factor of $\U(1)_R \times \SU(2)_F \subset \SU(2) \times \SU(2)_F \simeq \SO(4) \subset \SO(5)_R$. Before the twisting the supercharges transform under $\SO(3,1) \times \SO(2)_s \times \SU(2)_F \times \U(1)_R$ as:
\be
\rep{4} \otimes \rep{4} \quad\to\quad \big( \rep{2}_\frac{1}{2} + \rep{2}'_{-\frac1 2} \big) \otimes \big( \rep{2}_0 + \rep{1}_{\frac1 2} + \rep{1}_{-\frac1 2} \big) \;,
\ee
with symplectic Majorana condition. We twist $\SO(2)_s \to \SO(2)_s - \U(1)_R$. The supercharges transform as:
\be
\rep{2}_{\frac1 2} \otimes \rep{2}_0 + \rep{2}_0 \otimes \rep{1}_{\frac1 2} + \rep{2}_1 \otimes \rep{1}_{-\frac1 2} + \rep{2}'_{-\frac1 2} \otimes \rep{2}_0 + \rep{2}'_{-1} \otimes \rep{1}_{\frac1 2} + \rep{2}'_0 \otimes \rep{1}_{-\frac1 2} \;.
\ee
The preserved supercharges are $\rep{2}_0 \otimes \rep{1}_{\frac1 2}$ (with conjugate $\rep{2}'_0 \otimes \rep{1}_{-\frac1 2} = Q^\dag$), which give \Nugual{1}. $R_{\Nugual{1}}$ is identified with twice the charge under $\U(1)_R$. The scalars decompose as:
\be
\rep{1} \otimes \rep{5} \quad\to\quad \rep{1}_0 \otimes \big( \rep{2}_{\frac1 2} + \rep{2}_{-\frac1 2} + \rep{1}_0 \big) \quad\xrightarrow{\text{twisting}}\quad \rep{1}_{-\frac1 2} \otimes \rep{2}_{\frac1 2} + \rep{1}_{\frac1 2} \otimes \rep{2}_{-\frac1 2} + \rep{1}_0 \otimes \rep{1}_0 \;,
\ee
with the reality condition reducing to a relation between the first two terms on the right hand side. $\rep{1}_{-\frac1 2} \otimes \rep{2}_{\frac1 2}$ is a complex scalar, a spinor on the Riemann surface, doublet of $\SU(2)_F$, and has $R = 1$. $\rep{1}_0 \otimes \rep{1}_0$ is a real scalar, scalar on $\Sigma_g$ and has $R = 0$. It pairs up with $B_{ab}$ to form the complex scalar of a neutral chiral multiplet. The fermions $\rep{4}' \otimes \rep{4}$ give $\rep{2}_{-\frac12} \otimes \rep{2}_0 + \rep{2}_{-1} \otimes \rep{1}_{\frac12} + \rep{2}_{0} \otimes \rep{1}_{-\frac12}$ and conjugates.


\section{Explicit form of Maldacena-Nu\~nez solutions}
\label{app: MN metric}

The metric is of the following form:
\begin{multline}
ds_{11}^2= (\pi N l_p^3)^{2/3} \, \bigg[
\Delta^{1/3} ds_7^2 \\
+ \frac14 \, \Delta^{-2/3} \Big( X_0^{-1} d\mu_0^2 + \sum_{i=1,2} X_i^{-1} \, \big[ d\mu_i^2+\mu_i^2 \, (d\phi_i+ 2A_i)^2 \big] \Big)  \bigg] \;,
\end{multline}
where $l_p$ is the 11d Planck length,
\begin{equation}
\label{metric}
ds_7^2 = e^{2f} \, \frac{dr^2+dx^\mu dx_\mu}{r^2} + e^{2h} \, \frac{dx^2+dy^2}{y^2}
\end{equation}
is the metric of $AdS_5 \times H^2$, $A_i$ are 1-forms on $H^2$ and
\begin{equation}
\Delta = \sum_{i=0,1,2} X_i \, \mu_i^2 \;.
\end{equation}
The redundant coordinates $\mu_{0,1,2}$ are constrained by
\begin{equation}
\mu_0^2 + \mu_1^2 + \mu_2^2 = 1 \;,
\end{equation}
and parameterize a two-sphere. We parameterize $X_{0,1,2}$ via
\begin{equation}
X_0 = e^{-4(\lambda_1+\lambda_2)} \;, \qquad\qquad
X_1 = e^{2\lambda_1} \;, \qquad\qquad
X_2 = e^{2\lambda_2} \;.
\end{equation}
We refer the reader to the original paper \cite{Maldacena:2000mw}
for the form of the $G_4$ field.%
\footnote{Note that $h$ is called $g$ in the original paper; we renamed it
to avoid the confusion with the genus of the Riemann surface.}

The $\cN=2$ solution is given by
\begin{equation}
e^{2\lambda_1} = 2^{-3/5} \;, \qquad
e^{2\lambda_2} = 2^{2/5} \;, \qquad
e^{2f} = 2^{-4/5} \;, \qquad
e^{2h} = 2^{-9/5}
\end{equation}
and
\begin{equation}
A_1 = 0 \;, \qquad\qquad A_2 = \frac12 \, \frac{dx}y  \;.
\end{equation}
To exhibit the $\SU(2)_R \times \U(1)_R$ isometry, we perform the change of coordinates: $\mu_0 = \cos\theta\, \cos\psi$, $\mu_1 = \cos\theta\, \sin\psi$, $\mu_2 = \sin\theta$. Then $\Delta$ and $\mu_2$ are functions of $\theta$ only; $(\psi, \phi_1)$ parameterize a round $S^2$ on which $\SU(2)_R$ acts; $\U(1)_R$ acts as translations of $\phi_2$, which is the only direction to be fibred over $H^2$.

The $\cN=1$ solution is given by
\begin{equation}
e^{2\lambda_1} = e^{2\lambda_2} = \Big( \frac34 \Big)^{1/5} \;, \qquad
e^{2f} = \Big( \frac34 \Big)^{8/5} \;, \qquad
e^{2h} = \frac14 \, \Big( \frac34 \Big)^{3/5}
\end{equation}
and
\begin{equation}
A_1 = A_2 = \frac14 \, \frac{dx}y \;.
\end{equation}
Now $\Delta$ is a function of $\mu_0$ only. At fixed $\mu_0$ and $x$, the metric in parenthesis is a round $S^3$ of isometry $\SU(2) \times \SU(2)_F$. Both $d\mu_0^2$ and the fibration by $A_1 = A_2$ break it to $\U(1)_R \times \SU(2)_F$.

\bibliographystyle{utphys}
\bibliography{bib}{}

\end{document}